\documentclass[lettersize,journal]{IEEEtran}
\usepackage{amsmath,amsfonts}
\usepackage{algorithm}
\usepackage{array}
\usepackage{textcomp}
\usepackage{stfloats}
\usepackage{url}
\usepackage{verbatim}
\usepackage{graphicx}
\usepackage{wrapfig}
\usepackage{tcolorbox}
\usepackage{amssymb}
\usepackage{subcaption}
\usepackage{threeparttable}
\usepackage{booktabs}
\usepackage{cite}

\usepackage{setspace}
\usepackage[utf8]{inputenc} % allow utf-8 input
\usepackage[T1]{fontenc}    % use 8-bit T1 fonts
\usepackage{hyperref}       % hyperlinks
\usepackage{url}            % simple URL typesetting
\usepackage{booktabs}       % professional-quality tables
\usepackage{amsfonts}       % blackboard math symbols
\usepackage{nicefrac}       % compact symbols for 1/2, etc.
\usepackage{microtype}      % microtypography
\usepackage{xcolor}         % colors
\usepackage{wrapfig}
\usepackage{threeparttable}

\usepackage{amsmath,amssymb,amsfonts}
\usepackage{amssymb}
\usepackage{breqn}
\usepackage{ragged2e}
\usepackage{cleveref}
\usepackage{float}   

\usepackage{booktabs}
\usepackage{cases}
\usepackage{graphicx}
\usepackage{multirow}
\usepackage{mathrsfs}
\usepackage{textcomp}
\usepackage{xcolor}
\usepackage{bm}
\usepackage[noend]{algpseudocode}
\usepackage{tcolorbox}
\usepackage{algorithmicx,algorithm}
\usepackage{subcaption}
\usepackage{amssymb}
\usepackage{graphicx}
\usepackage{amsmath}
\usepackage{braket}
\usepackage{gensymb}
\usepackage{multirow}
\usepackage{booktabs}
\usepackage{adjustbox}
\usepackage{array}
\usepackage{bm}
\usepackage{mciteplus}
\usepackage{colortbl}
\usepackage{xcolor}
\definecolor{myred}{rgb}{1, 0.698, 0.698}
\definecolor{myorange}{rgb}{1,0.851,0.698}
\definecolor{myyellow}{rgb}{1, 0.973, 0.773}
\usepackage{booktabs} % Because you are using commands like \toprule
\usepackage{amsmath}  % Because you are using the \downarrow symbol
\usepackage{multirow} % Added for the \multirow command
\usepackage[table]{xcolor}   % <-- 添加了 [table] 选项以支持 \cellcolor
\usepackage[table]{xcolor}   % <-- 添加了 [table] 选项以支持 \cellcolor
\usepackage{makecell} 

\usepackage{colortbl}
\usepackage[table]{xcolor}
\usepackage[table]{xcolor} % 导入xcolor包，并启用table选项

\definecolor{MyBlue1}{HTML}{BCC6DD} % 最浅 (对应差结果)
\definecolor{MyBlue2}{HTML}{98A3CA}
\definecolor{MyBlue3}{HTML}{8092C4}
\definecolor{MyBlue4}{HTML}{455D99} % 最深 (对应好结果)

\usepackage[table]{xcolor} % 确保已加载 xcolor 包
\usepackage{hhline}       % 导入 hhline 包, 解决线条和颜色的冲突

\definecolor{MyRed1}{HTML}{D0DCE8} % 最浅 (对应最差结果, 224.37)
\definecolor{MyRed2}{HTML}{BFD0E1} % 
\definecolor{MyRed3}{HTML}{B5CBE2} % 
\definecolor{MyRed4}{HTML}{94B5D7} % 

\definecolor{MyRed5}{HTML}{8092c4} % 最深 (对应最好结果)

\definecolor{MyGreen1}{HTML}{FBE9EA} % 最浅 (对应差结果 ~10x)
\definecolor{MyGreen4}{HTML}{EBA1A8}% 最深 (对应好结果 ~100x)
\begin{document}

\title{Geometry Modality-Aware Multimodal Alignment for Geometry-Free Fast Molecular Quantum Hamiltonian Prediction}

\title{Unlocking Geometry Perception in Molecular Language via Modality Compensation for Fast Molecular Quantum Hamiltonian Prediction}

\title{Endowing Molecular Language with Geometry Perception via Modality Compensation for High-Throughput Quantum Hamiltonian Prediction}

\author{
Zhenzhong Wang,~\IEEEmembership{Member, IEEE},
Yongjie Hou,
Chenggong Huang,
Yuxuan Du,\\
Dacheng Tao,~\IEEEmembership{Fellow, IEEE},
Min Jiang\textsuperscript{*},~\IEEEmembership{Senior Member, IEEE}
 % <-this % stops a space
\IEEEcompsocitemizethanks{\IEEEcompsocthanksitem
 % This work was supported in part by the National Natural Science Foundation of China under Grant U21A20512; in part by the National Natural Science Foundation of China under Grant 62276222; in part by the Research Grants Council of the Hong Kong under Grant PolyU11211521 and Grant PolyU15218622; in part by the Hong Kong Polytechnic University (Project No.: 1-ZE0C). ().

Zhenzhong Wang, Chenggong Huang, and Min Jiang are with the Department of Artificial Intelligence, Key Laboratory of Multimedia Trusted Perception and Efficient Computing, Ministry of Education of China, Key Laboratory of Digital Protection and Intelligent Processing of Intangible Cultural Heritage of Fujian and Taiwan, Ministry of Culture and Tourism, School of Informatics, Xiamen University, Xiamen 361005, Fujian, P.R. China  (e-mail: zhenzhongwang@xmu.edu.cn; huangchenggong@stu.xmu.edu.cn; minjiang@xmu.edu.cn).

Yongjie Hou is with the School of Electronic Science and Engineering, Xiamen University, Xiamen 361005, Fujian, P.R. China  (e-mail: 23120231150268@stu.xmu.edu.cn). 

Yuxuan Du is with the College of Computing and
Data Science and School of Physical and Mathematical Science, Nanyang Technological University, Singapore 639798 (e-mail: duyuxuan123@gmail.com).

Dacheng Tao is with the College of Computing and
Data Science, Nanyang Technological University, Singapore 639798 (e-mail: dacheng.tao@gmail.com).

\textit{*Corresponding author: Min Jiang.}

% This work was supported in part by the National Natural Science Foundation of China under Grants No. U21A20512; in part by the Research Grants Council of the Hong Kong SAR under Grant C5052-23G, Grant PolyU 11211521, Grant PolyU 15218622, and Grant PolyU 15215623; in part by the Hong Kong Research Grant Council General Research Fund under Grant PolyU 15208222; in part by the National Natural Science Foundation of China Young Scientist Fund under Grant PolyU A0040473 and Start-Up Fund for New Recruits under Grant PolyU A0046682; \textit{\textsuperscript{†}These authors contributed equally to this work.} \textit{(*Corresponding author: Wanyu Lin.)} 

		% note need leading \protect in front of \\ to get a newline within \thanks as
		% \\ is fragile and will error, could use \hfil\break instead.

	}	
}

% The paper headers
% \markboth{Journal of \LaTeX\ Class Files,~Vol.~14, No.~8, August~2021}%
% {Shell \MakeLowercase{\textit{et al.}}: A Sample Article Using IEEEtran.cls for IEEE Journals}

% \IEEEpubid{0000--0000/00\$00.00~\copyright~2021 IEEE}
% Remember, if you use this you must call \IEEEpubidadjcol in the second
% column for its text to clear the IEEEpubid mark.

\maketitle

\begin{abstract}

The quantum Hamiltonian is a fundamental property that governs a molecule's electronic structure and behavior, and its calculation and prediction are paramount in computational chemistry and materials science. Accurate prediction is highly reliant on extensive training data, including precise molecular geometries and the Hamiltonian matrices, which are expensive to acquire via either experimental or computational methods. Towards a fast yet accurate method for Hamiltonian prediction, we first introduce a geometry information-aware molecular language model to bypass the use of expensive molecular geometries by only using the readily available molecular language --- simplified molecular input line entry system (SMILES). Our method employs multimodal alignment to bridge the relationship between SMILES strings and their corresponding molecular geometries. Recognizing that the molecular language inherently lacks explicit geometric information, we propose a geometry modality compensation strategy to imbue molecular language representations with essential geometric features, thereby enabling accurate predictions using SMILES. 
In addition, given the high cost of acquiring Hamiltonian data, we devise a weakly supervised strategy to fine-tune the molecular language model, thus improving the data efficiency.
Theoretically, we prove that the prediction generalization error without explicit molecular geometry can be bounded through our modality compensation scheme.
Empirically, our method achieves superior computational efficiency, providing up to a $\sim$100.0$\times$ speedup over conventional quantum mechanical methods while maintaining comparable prediction accuracy.
We further demonstrate the practical case study of our approach in the screening of electrolyte formulations. 
The extensive experimental validations confirm its promise as a powerful tool for AI-based high-throughput quantum chemistry. 

% The code is available at \url{https://anonymous.4open.science/r/MGAHam-966F}. 
\end{abstract}


\section{Introduction}

\begin{figure*}[htbp]
	\centering
	\includegraphics[width=0.98\linewidth]{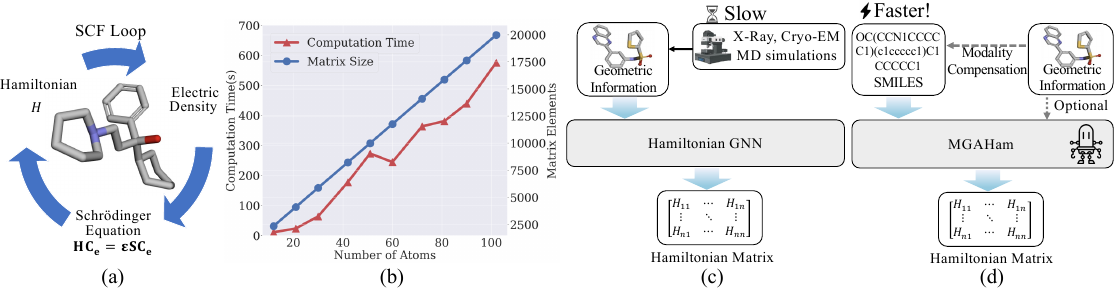}
	\caption{(a) DFT calculates the Hamiltonian by solving the Schrödinger equation based on electron density. (b) As the molecular size grows, both the dimensionality of the Hamiltonian matrix and the associated computational cost of DFT increase significantly. (c) Conventional Hamiltonian GNNs require precise molecular geometries, the acquisition of which necessitates costly experimental or computational methods. (d) The proposed {\em MGAHam} enables rapid Hamiltonian prediction by circumventing the need for 3D geometry, thereby achieving substantial computational speedups.}
	\label{fig:intro}
        % \vspace{-0.1cm}
\end{figure*}

The quantum Hamiltonian fundamentally reflects a molecule's electron structures and energy states, consequently governing almost all electronic properties, including charge density, band structure, and response to electromagnetic fields~\cite{wang2024crystalline,gong2023general,gu2024deep,NEURIPS2023_7f755e27}. 
Density Functional Theory (DFT) has long been a widely adopted framework for practical Hamiltonian construction in electronic structure calculations~\cite{dft}. 
As shown in Fig. \ref{fig:intro} (a) and (b), DFT simplifies the complex, many-body electron problem into a more manageable system of non-interacting electrons, which is then solved via the Schrödinger equation. Despite its widespread use, the iterative self-consistent field (SCF) procedure suffers from computational costs that scale steeply with system size, rendering it prohibitively expensive for large-scale systems.

The computational bottleneck of traditional approaches has motivated the development of AI-based models~\cite{10595522,wu2024tamgen,10113742,geng2023de,11123751,10443285,zhang2024overcoming,10706014,zhang2024self,PEGNet,wang2025cross,10164235,chen2021data,yin2025advancing,Wong2025EvolutionaryOO}. Notably, geometric neural networks (GNNs)~\cite{wu2020comprehensive,wang2024crystalline} leverage message-passing mechanisms to model atomic interactions explicitly.
Among existing frameworks, SchNOrb~\cite{schutt2019unifying} utilizes pairwise distances to estimate Hamiltonian matrices but lacks inherent equivariance. While PhiSNet~\cite{unke2021se} ensures equivariance through specialized architectures, it is constrained by fixed-size molecule inputs. DeepH~\cite{li2022deepDFT} provides high accuracy through local coordinate systems but is primarily optimized for crystalline materials. Recent advancements like QHNet~\cite{yu2023efficient} and its successor DEQHNet~\cite{wang2024infusing} offer greater versatility and improved electron density descriptions by introducing expandable modules and density-equivariant schemes. 

% However, these methods generally rely on explicit 3D geometric inputs, posing challenges for scenarios where 3D configurations are unavailable or costly to obtain.

While these GNNs have made remarkable advances, their fundamental reliance on precise molecular geometries as input poses a significant bottleneck~\cite{fogarasi1992calculation}. As shown in Fig. \ref{fig:intro} (c), obtaining these geometries through either experimental techniques (e.g., X-ray diffraction~\cite{drenth2007principles} and Cryo-EM~\cite{danev2019cryo}) or computational methods (e.g., molecular dynamics simulations~\cite{badar2022molecular}) remains prohibitively expensive. This impedes their application in scenarios where accurate geometries are unavailable, hindering the potential for high-throughput screening of Hamiltonians and related properties, such as the highest occupied molecular orbital (HOMO), the lowest unoccupied molecular orbital (LUMO), and the HOMO-LUMO gap~\cite{zheng2023predicting}, all of which are critical to understanding chemical reactivity. Therefore, developing data-efficient models is paramount for accelerating early-stage molecular discovery, as they enable robust training and accurate predictions even when initial data is highly limited.

%,omar2021high,choi2022scalable,abroshan2023high}.

% In contrast to molecular geometries, molecular languages such as SMILES~\cite{weininger1988smiles} are highly accessible. Extensive studies have demonstrated the feasibility of utilizing large language models for various molecular property predictions, including toxicity. 

Molecular languages like SMILES~\cite{weininger1988smiles} are far more accessible than precise 3D geometries, and their utility has largely been focused on predicting macroscopic properties such as toxicity or solubility~\cite{zheng2025large,wang2024explainable,pinheiro2020machine,xu2023transpolymer}. These properties are often governed by specific molecular fragments rather than precise atomic coordinates. In contrast, utilizing molecular languages to predict microscopic properties remains challenging and limited. This research gap reflects a fundamental imbalance in representational capacity. Molecular languages function as an information-deficient modality, as they abstract away the 3D spatial context necessary for capturing interatomic interactions. Conversely, molecular geometries serve as an information-rich modality, encapsulating precise spatial arrangements.

To exploit the synergy between different modalities, multimodal learning has emerged as a dominant paradigm. They typically bridge disparate modalities by establishing cross-modal associations within a shared latent space \cite{liu2023molca,11172339,litowards,11192222,liunext,ock2024multimodal,wu2024asymmetric}. For instance, models can leverage an information-deficient modality (e.g., text) to guide the synthesis or retrieval of an information-rich modality (e.g., images). However, how to independently leverage data from an information-deficient modality (e.g., molecular language) to achieve the representational power of an information-rich modality (e.g., molecular geometry) remains a fundamental open problem in multimodal learning.

% In addition, existing deep learning approaches typically require large-scale training datasets. This requirement poses a challenge in scenarios such as characterizing molecular behaviors under specific experimental conditions, where both molecular geometries and Hamiltonian data are often scarce. Consequently, 

In this work, we propose {\em MGAHam}, a novel \underline{m}ultimodal language model equipped with a \underline{g}eometry-\underline{a}ware mechanism for fast yet accurate molecular \underline{Ham}iltonian prediction. Notably, {\em MGAHam} supports inference using either full molecular geometries or solely SMILES strings (see Fig. \ref{fig:intro} (d)). To bypass the requirement for expensive molecular geometries during inference, we first establish the correspondence between SMILES strings and their associated molecular geometries through a multimodal pre-training strategy. Recognizing the critical role of local atomic environments in Hamiltonian calculation~\cite{li2022deepDFT}, we employ a local-environment-aware alignment method to link molecular language fragments with their respective local geometric counterparts. Furthermore, to empower the information-deficient SMILES with the representational capacity of the information-rich molecular geometry, we introduce a learnable affine transformation-based modality compensation strategy. This strategy transfers essential spatial information into the SMILES embeddings, thereby enhancing their geometric awareness. In addition, to address the high cost of acquiring Hamiltonian matrices, we devise a mask-based weakly supervised strategy during the fine-tuning stage. Our key contributions are summarized as follows:

\begin{table}[t]
\centering
\footnotesize 
\setlength{\tabcolsep}{4pt} 
% --- --- ---
\caption{The summarized comparison results of the proposed {\em MGAHam} against baselines on the $\text{QH}$ series datasets.}
\label{tab:model_comparison}
% --- Modified Table ---
\resizebox{0.99\linewidth}{!}{
\begin{tabular}{cccc}
\toprule
\makecell{Model} & \makecell{Input \\Modality} & \makecell{Acceleration \\Ratio to DFT} & \makecell{Total MAE [$E_h$] on all elements \\of Hamiltonian matrices.} \\
\midrule
% SMILES-BERT & 1D String & \cellcolor{MyRed4}$\sim$100$\times$ & \cellcolor{MyGreen1}$\sim 6.0\times10^{-2}- 3.1\times10^{-1}$  \\
% PhiSNet & 3D Geometry & \cellcolor{MyGreen1}$\sim$10$\times$ & \cellcolor{MyRed1}$\sim 6.1\times10^{-2} - 2.2\times10^{-1}$ \\
QHNet~\cite{yu2023efficient} & 3D Geometry & \cellcolor{MyRed1}$\sim$10$\times$ & \cellcolor{MyGreen1}$\sim7.0\times10^{-5}- 3.0\times10^{-1} $ \\
% SchNOrb & 3D Geometry & \cellcolor{MyGreen1}$\sim$10$\times$ & \cellcolor{MyRed3}$\sim 1.1\times10^{-2} - 4.2\times10^{-1}$ \\
DEQHNet~\cite{wang2024infusing} & 3D Geometry & \cellcolor{MyRed1}$\sim$10$\times$ & \cellcolor{MyGreen1}$\sim 7.0\times10^{-5}-3.0\times10^{-1}  $ \\
SE(3)-Transformer\cite{fuchs2020se} & 3D Geometry & \cellcolor{MyRed1}$\sim$10$\times$ & \cellcolor{MyGreen1}$\sim 8.0\times10^{-5}- 3.0\times10^{-1} $ \\
GemNet\cite{gasteiger2021gemnet} & 3D Geometry & \cellcolor{MyRed1}$\sim$10$\times$ & \cellcolor{MyGreen1}$\sim 7.0\times10^{-5}-3.0\times10^{-1}  $ \\
\cmidrule(lr){1-4} 
\multirow{2}{*}{{\em MGAHam}} & 1D String & \cellcolor{MyRed4}$\sim$100$\times$ & \cellcolor{MyGreen4}$\sim 7.0\times10^{-5}-2.0\times10^{-1} $ \\ 
\hhline{~---}         
                     & 1D String + 3D Geometry & \cellcolor{MyRed1}$\sim$10$\times$ & \cellcolor{MyGreen4}$\sim 7.0\times10^{-5}-2.0\times10^{-1} $ \\ 
\bottomrule
\end{tabular}}
% \vspace{-0.3cm}
\end{table}

\begin{enumerate}
    \item We introduce a novel paradigm for deep learning-based Hamiltonian prediction. Our proposed {\em MGAHam} incorporates a fine-grained multimodal alignment that establishes a precise correspondence between molecular language fragments and their local atomic environments. The proposed multimodal method enables the model to accept SMILES strings as input, eliminating the dependency on costly molecular geometries during inference. Our theoretical analysis demonstrates that the generalization error arising from the absence of explicit geometric data can be upper-bounded within the proposed framework.

    % error bound is upper bounded by some factors, e.g., # trainable params, # training examples, Gradients norm, etc
    
    % \item To enable molecular language representations to serve as a surrogate for molecular geometries, 
    
    \item Recognizing that molecular languages inherently lack explicit geometric information, we propose a geometry modality compensation strategy to effectively transfer essential geometric features into the SMILES embeddings, thereby enabling accurate predictions in the absence of explicit atomic coordinates without sacrificing performance. 
    
    \item To address the high cost of Hamiltonian acquisition, we introduce a weakly supervised strategy that leverages limited and incomplete Hamiltonian information as training data, which enhances the model's generalizability in low-data regimes. 
    
    \item Extensive evaluations on four datasets---including MD17, QH9, our curated large-molecule dataset (QH-BM), and the high-temperature dataset (QH9-1000K)---demonstrate that {\em MGAHam} achieves promising performance. As shown in Table \ref{tab:model_comparison}, {\em MGAHam} offers a speedup of approximately 100$\times$ over DFT while maintaining prediction accuracy comparable to geometry-dependent GNNs, even when both Hamiltonian and geometric data are scarce. This approach establishes a promising new avenue for AI-driven high-throughput quantum chemistry.
\end{enumerate}

\section{Preliminaries}

\subsection{DFT Hamiltonian}
DFT is a widely used framework for modeling the electronic structure and ground-state properties of many-electron systems\cite{Hohenberg64, dft}. Its fundamental basis lies in the Hohenberg-Kohn theorems, which recast the intractable many-electron problem into a computationally tractable auxiliary system of non-interacting electrons. This fictitious system is governed by the Kohn-Sham equations, where electrons move within an effective potential that accounts for the external potential from the nuclei, the classical Hartree repulsion, and quantum mechanical exchange-correlation effects. To determine the ground-state properties, the one-electron Kohn-Sham molecular orbitals, $\psi_i(\bm{r})$, are conventionally approximated as a linear combination of atomic orbitals using a predefined set of basis functions, $\{\phi_j(\bm{r})\}$:
\begin{equation}
\psi_i(\bm{r}) = \sum_j C_{ij}\phi_j(\bm{r}),
\end{equation}
where $C_{ij}$ denotes the molecular orbital expansion coefficients, which represent the contribution of each atomic basis function $\phi_j(\bm{r})$ to the $i$-th molecular orbital $\psi_i(\bm{r})$.

This discretization converts the continuous problem into a generalized matrix eigenvalue equation \cite{szabo1996modern}:
\begin{equation}\label{eq:HCSC}
 \mathbf{H}\mathbf{C} = \mathbf{S}\mathbf{C}\epsilon.
\end{equation}
In this formulation, $\mathbf{H}$ is the Kohn-Sham matrix representing the effective kinetic and potential energy terms, $\mathbf{S}$ is the overlap matrix of the basis functions, $\mathbf{C}=\{C_{ij}\}$ contains the molecular orbital coefficients, and $\epsilon$ is the diagonal matrix of orbital energies. The solution is obtained iteratively through the SCF procedure \cite{payne1992iterative}. Crucially, the computational expense of the SCF cycle scales with respect to the system size. This scaling renders DFT computationally prohibitive for large-scale molecular simulations, thereby motivating the necessity for more efficient alternatives.

\subsection{Hamiltonian GNNs}

To overcome the computational bottlenecks of traditional methods such as DFT, geometric neural networks (GNNs) have rapidly emerged as a powerful learning paradigm, offering a compelling pathway to accelerated calculations \cite{liu2021spherical, se3transformers, schutt2021painn, godwin2021noisynode, satorras2021egnn, qiao2020orbnet}. By representing molecules as graphs, these models enable the learning of complex mappings from atomic coordinates to Hamiltonian matrices, thus serving as computationally efficient DFT surrogates.

 Initial methods, exemplified by SchNOrb \cite{schutt2019unifying}, adapted distance-based convolution by integrating symmetry-adapted atomic orbital features and incorporating pairwise representations. While effective, these early models did not enforce strict equivariance by construction. Subsequent research focused intensely on building explicitly $\text{E}(3)$-equivariant architectures. PhiSNet \cite{unke2021se} marked a significant breakthrough by introducing a systematic method to construct equivariant features directly from atomic coordinates. Following this foundation, QHNet \cite{yu2023efficient} further refined the equivariant design through innovations such as a learnable norm gate and node-wise attention, boosting both accuracy and computational speed. More recent work, such as DEQHNet \cite{wang2024infusing}, has deepened the physical fidelity by incorporating a density-equivariant message passing scheme, thereby ensuring that features related to the electron density maintain the correct transformation properties. This clear technical progression highlights a sustained commitment within the field to developing more principled and physically grounded equivariant models for efficient electronic structure prediction. 
 
Despite these achievements, their inherent reliance on extensive molecular geometries and the Hamiltonian matrices fundamentally restricts their utility in scenarios, such as high-throughput screening, where accurate molecular geometries may be unavailable. In this work, we investigate the intrinsic correlation between the information-rich molecular geometry and the information-deficient molecular language. By bridging this gap, we endow molecular languages with the representational capacity equivalent to molecular geometries, thereby bypassing the use of costly 3D geometries.

% Pioneering methods, such as SchNOrb \cite{schutt2019unifying} and PhiSNet \cite{unke2021se}, initially focused on constructing the Hamiltonian by employing symmetry-adapted and equivariant representations of molecular structures. Building on this foundation, more advanced architectures have been developed. For instance, QHNet \cite{yu2023efficient} introduced a specialized design that uses a learnable norm gate and attention mechanisms to achieve high accuracy with minimal memory overhead. Furthermore, DEQHNet \cite{wang2024infusing} incorporated a density-equivariant message passing scheme, enhancing the model's ability to handle information related to electron density distribution.

\section{Methodology}

\subsection{Framework}

\begin{figure*}[t]
	\centering
	\includegraphics[width=0.998\linewidth]{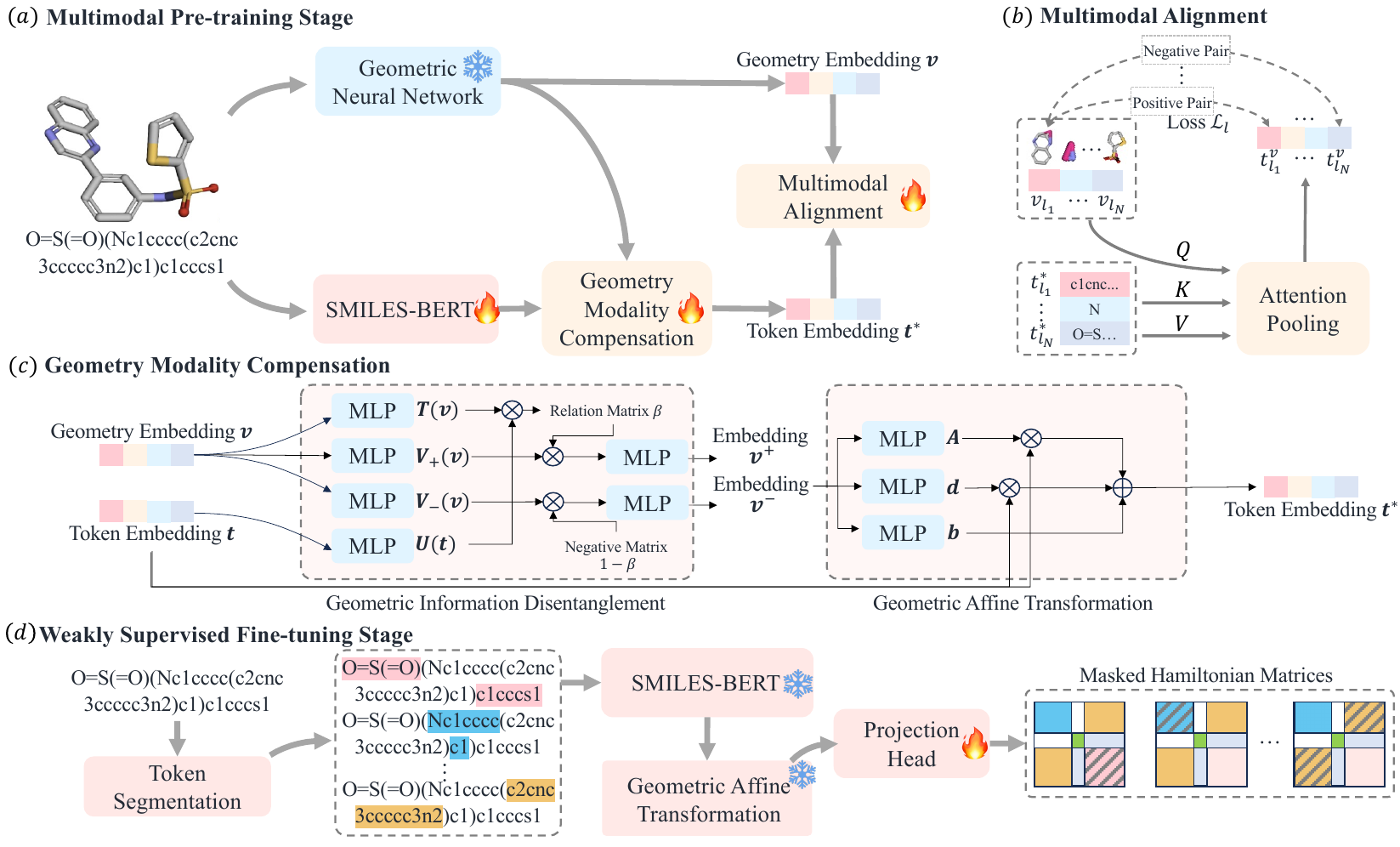}
	\caption{The pipeline of the proposed {\em MGAHam} framework. (a) {\em MGAHam} processes SMILES strings via SMILES-BERT to generate token embeddings $\mathbf{t}$, while molecular geometries are processed by QHNet to produce geometric embeddings $\mathbf{v}$. The token embeddings $\mathbf{t}$ then undergo modality compensation to generate geometry-aware token embeddings $\mathbf{t}^*$ for multimodal alignment. (b) A cross-modal projector aligns the token and geometric embeddings into a shared embedding space. (c) The modality compensation module transfers geometric information from geometric embeddings $\mathbf{v}$ to token embeddings $\mathbf{t}$, producing the geometry-aware token embeddings $\mathbf{t}^*$. (d) During the weakly supervised fine-tuning stage, the model operates exclusively on SMILES strings, enabling rapid and accurate Hamiltonian prediction without the need for molecular geometries.}
	\label{pipeline}
    % \vspace{-0.2cm}
\end{figure*}

To establish the relationship between molecular language and molecular geometry, we employ a multimodal learning paradigm to align these two modalities within a shared latent space. In particular, we design a fine-grained multimodal alignment and a geometric modality compensation strategy, endowing molecular language with geometry perception.

Fig. \ref{pipeline} illustrates the pipeline of the proposed {\em MGAHam}. Firstly, the multimodal pre-training framework is designed to bridge the gap between 1D SMILES strings and 3D molecular geometries by establishing a shared latent space. As depicted in Fig. \ref{pipeline} (a), the process begins by encoding SMILES inputs into token embeddings using SMILES-BERT, while a parallel GNN architecture generates geometric embeddings from corresponding 3D structures. To address the inherent information asymmetry between these modalities, we employ a geometry compensation module that transfers critical geometric knowledge to the token embeddings. This is followed by a local atomic environment-aware alignment module, which performs fine-grained matching between specific SMILES fragments and their local geometric contexts. By integrating these strategies, our pre-training scheme imbues the token embeddings with essential geometric features, enabling rapid and accurate Hamiltonian prediction without relying on explicit 3D molecular geometries during inference.

Following the multimodal pre-training, the subsequent fine-tuning stage requires only SMILES string inputs to establish the mapping to Hamiltonian matrices. To address the high cost of data acquisition, this stage utilizes a mask-based weakly supervised strategy to enhance the model's generalization performance in low-data regimes. Details of the {\em MGAHam} architecture and its implementation are provided in what follows.

% \begin{figure*}[t]
% 	\centering
% 	\includegraphics[width=0.998\linewidth]{pictures/pipeline-yj.pdf}
% 	\caption{The pipeline of our proposed {\em MGAHam}. (a) {\em MGAHam} processes SMILES strings using SMILES-BERT to generate token embeddings $\mathbf{t}$ and molecular geometries using QHNet to produce geometric embeddings $\mathbf{v}$, and the token embeddings $\mathbf{t}$ go through the asymmetric modality compensation to generate geometry-aware token embeddings $\mathbf{t}^*$ for the multimodal alignment. (b) A cross-modal projector aligns the token embeddings and geometric embeddings into a shared embedding space. (c) The asymmetric modality compensation module transfers geometric information from geometric embeddings $\mathbf{v}$ to token embeddings $\mathbf{t}$ to produce the geometry-aware token embeddings $\mathbf{t}^*$. (d) The weakly supervised fine-tuning strategy operates exclusively on the SMILES strings, enabling rapid yet accurate Hamiltonian prediction.}
% 	\label{pipeline}
% \end{figure*}

\subsection{Multimodal Pre-training Stage}

Multimodal pre-training aims to establish the correspondence between SMILES strings and their geometric counterparts via latent space alignment. As depicted in Fig. \ref{pipeline} (a), this process begins with SMILES-BERT\cite{wang2019smiles} extracting token embeddings from SMILES inputs, while QHNet\cite{yu2023efficient} generates geometric embeddings from geometric structures. \footnote{We also investigate using different language models and geometric neural networks as encoders to extract token embeddings and geometric embeddings, respectively. The impact of using different encoders can be found in the Supplemental Material.} Subsequently, two core modules are incorporated: (1) a geometry modality compensation module that transfers geometric knowledge from the geometric embeddings to enrich the information-poorer token embeddings; and (2) a local atomic environment aware multimodal alignment module that enables fine-grained local matching between SMILES fragments and their corresponding local geometric atomic environments. Through these strategies, multimodal pre-training imbues token embeddings with essential geometric features, thereby ensuring efficient and accurate Hamiltonian prediction even when explicit molecular geometries are bypassed.

% \textbf{Geometric Information Disentanglement.}

\textbf{Geometry Modality Compensation.} Traditional multimodal pre-training methods typically align different modalities directly, often overlooking their inherent information discrepancies. Consequently, modalities with weak discriminative information could limit the model's overall prediction capability~\cite{li2021asymmetric,wu2024asymmetric}. This limitation becomes particularly detrimental in Hamiltonian prediction when only using the data of weak modal --- SMILES strings. Therefore, we propose to transfer geometric knowledge from geometric embedding to the token embedding to enrich it with geometry-aware capabilities.

As presented in Fig. \ref{pipeline} (c), we first disentangle geometric embedding $\mathbf{v}$, as the output of QHNet, into token embedding-relevant and token embedding-irrelevant components. The token embedding-irrelevant component retains essential geometric information and is used to enhance the token embedding, ensuring geometric awareness for SMILES inputs. 
The intuition behind information disentanglement is to identify the component from the geometric embedding that is complementary to the token embedding, i.e., token embedding-irrelevant. This complementary information is then transferred to the token embedding to enhance its representation capability.
Specifically, we use cross-modal attention to disentangle the geometric embedding $\mathbf{v}\in \mathbb{R}^N$ into token embedding-relevant features and token embedding-irrelevant features. 
The cross-modal attention matrix $\bm{\beta} \in [0,1]^{N \times N}$ describes the correlation between geometric embedding and token embedding, and it can be computed as:
% Let $\mathbf{f}_s \in \mathbb{R}^{N \times D}$ and $\mathbf{f}_g \in \mathbb{R}^{N \times D}$ represent 1D sequential and 3D geometric features respectively, where $N$ is the number of atoms/tokens and $D$ is the feature dimension.
\begin{equation}
    \beta_{i,j} = \frac{\exp\left(\cos \langle U(\mathbf{v}_i), T(\mathbf{t}_j) \rangle\right)}{\sum_j \exp\left(\cos \langle U(\mathbf{v}_i), T(\mathbf{t}_j) \rangle\right)},
\end{equation}
where $U(\cdot)$ and $T(\cdot)$ are MLPs with parameters $W_u$ and $W_t$. From this correlation matrix, we can disentangle token embedding-irrelevant features $\mathbf{v}^-$ that encapsulate essential geometric information as well as token embedding-relevant features $\mathbf{v}^+$:
\begin{equation}
    \mathbf{v}^+ = \bm{\beta}V_+(\mathbf{v}),\quad \mathbf{v}^- = (\mathbf{I} - \bm{\beta})V_-(\mathbf{v}),
\end{equation}
where $V_+$ and $V_-$ are MLPs that map geometric embedding $\mathbf{v}$ to the token embedding-relevant representation space and the token embedding-irrelevant representation space. The token embedding-irrelevant features $\mathbf{v}^-$, encapsulating essential geometric information, are transferred into the token embedding to enrich its representation ability. 

%In this manner, the 1D-relevant feature $\mathbf{v}^+$ captures the consistency of the sequence topology, whereas the 1D-irrelevant feature $\mathbf{v}^-$ retains geometric information. 

% \textbf{Geometric Affine-based Information Transfer.} 

% While prior work has explored resolving modality asymmetry in multimodal models—for example, Sain et al. introduced a symmetric disentanglement model to decompose sketch and photo modalities into modality-shared content and modality-specific style components, then leveraging the shared content for cross-modality matching—these approaches are ill-suited for 1D-3D asymmetry. These existing methods primarily target 1D-2D asymmetry and fail to preserve the geometric equivariance (e.g., SO(3)-equivariance) required for Hamiltonian prediction.

Learnable affine transformations are widely employed for feature scaling and mapping to facilitate distribution alignment across disparate data domains \cite{bach2015batch}. Therefore, we design an affine transformation module to transfer the retained geometric information from token-irrelevant features $\mathbf{v}^-$ to the token embeddings $\mathbf{t}$, compensating for the inherent lack of geometric information in SMILES inputs. To capture the complex non-linear mapping between geometry information and token embeddings, the proposed geometric affine transformation incorporates non-linear terms. Specifically, the transformation is formulated as:
% learnable affine transformations~\cite{huang2017arbitrary}, 
\begin{equation}
\mathbf{t}^* = \mathbf{A}\mathbf{t} + \mathbf{b} + \mathbf{d}(\mathbf{t}),
\end{equation}
where $\mathbf{A} \in \mathbb{R}^{N \times N}$ is the transformation matrix, $\mathbf{b} = [b_1, b_2, \ldots, b_n]^\top \in \mathbb{R}^N$ is a simple translation vector, and $\mathbf{d}(\mathbf{t})\in \mathbb{R}^N$ is a nonlinear deformation vector. 

Since an affine transformation is a composite of rotation, scaling, and shearing, the transformation matrix $\mathbf{A}$ can be decomposed as:
\begin{equation}
\mathbf{A} =\mathbf{R}\cdot \mathbf{S}\cdot\mathbf{H},
\end{equation}
where $\mathbf{R}$ is the rotation matrix,  $\mathbf{S}$ is the scaling matrix, and $\mathbf{H}$ is the shear matrix. The Givens rotation matrix $\mathbf{R}_{ij}(\theta) \in \mathbb{R}^{N \times N}$ for rotation in the $(i,j)$-plane is defined with
\begin{equation}
\mathbf{R}_{ij}(\theta) = \begin{bmatrix}
1 & \cdots & 0 & \cdots & 0 & \cdots & 0 \\
\vdots & \ddots & \vdots & & \vdots & & \vdots \\
0 & \cdots & \cos\theta & \cdots & -\sin\theta & \cdots & 0 \\
\vdots & & \vdots & \ddots & \vdots & & \vdots \\
0 & \cdots & \sin\theta & \cdots & \cos\theta & \cdots & 0 \\
\vdots & & \vdots & & \vdots & \ddots & \vdots \\
0 & \cdots & 0 & \cdots & 0 & \cdots & 1
\end{bmatrix},
\end{equation}
where $\mathbf{R}_{ii} = \mathbf{R}_{jj} = \cos\theta$, $\mathbf{R}_{ij} = -\sin\theta$, $\mathbf{R}_{ji}= \sin\theta$, and all other entries match the identity matrix.

The scaling matrix $\mathbf{S} \in \mathbb{R}^{N \times N}$ is a diagonal matrix:
$\mathbf{S}  = \mathrm{diag}(s_1,\ldots,s_N)$ with learnable scaling factors $\{s_i\}_{i=1}^N$. The shear matrix $\mathbf{H} \in \mathbb{R}^{N \times N}$ takes the form:
\begin{equation}
\mathbf{H} = \mathbf{I} + \sum_{k=1}^K \mathbf{p}_k\mathbf{w}_k^\top,
\end{equation}
where $\mathbf{p}_k, \mathbf{w}_k \in \mathbb{R}^N$ are learnable shear direction vectors.
The nonlinear deformation $\mathbf{d}(\mathbf{t})$ uses high-dimensional sinusoidal perturbation:
\begin{equation}
\mathbf{d}(\mathbf{t}) = \begin{bmatrix}
a_1\sin(\omega_1 \mathbf{t}_1+\phi_1) \\
a_2\sin(\omega_2 \mathbf{t}_2+\phi_2) \\
\vdots \\
a_N\sin(\omega_N \mathbf{t}_N+\phi_N)
\end{bmatrix}.
\end{equation}
All parameters including rotation angles $\theta_{ij}$, scaling factors $s_i$, shear components $\mathbf{p}_k, \mathbf{w}_k$, translation components $b_i$, sinusoidal parameters $a_i$, $\omega_i$, $\phi_i$ are learned from $\mathbf{v}^-$ through MLPs.

To guarantee that the acquired geometry-aware token embedding $\mathbf{t}^*$ effectively capture molecular geometric information, we propose a modality discrepancy loss function $\mathcal{L}_D$ that jointly optimizes both modality representations:
\begin{equation}
\mathcal{L}^D = D(\mathbf{v}, {\mathbf{t}}^*) + \lambda_1 D(\mathbf{t}, \mathbf{v}^+),
\end{equation}
where $D(\cdot)$ denotes a distance metric that constrains the discrepancy between aligned cross-modality features. In our implementation, we employ the smooth L1 loss as the distance function. The hyperparameter $\lambda_1$ balances the weight between these two optimization objectives.

%ensuring neither modality dominates the shared embedding space.

% , while the influence of distant atoms is relatively minor. Consider a benzene molecule (C$_{6}$H$_{6}$) as an example. The local Hamiltonian of a single carbon-carbon (C-C) bond in the benzene ring is determined by the local atomic environment, which includes the two carbon atoms involved in the bond, their attached hydrogen atoms, and the adjacent carbon atoms in the ring. The delocalized $\pi$-electron system, formed by the overlapping $p$-orbitals of the carbon atoms, creates a unique electronic environment that influences the local Hamiltonian. In contrast, atoms or bonds farther away from this local region, such as those in a substituent group attached to the benzene ring, have a much smaller impact on the local Hamiltonian of the C-C bond. Therefore, learning the local chemical environment is crucial for predicting the Hamiltonian.

\textbf{Local Atomic Environment Aware Multimodal Alignment.} Local chemical environment affects the local electronic structure of molecules. 
Taking the benzene molecule (C$_{6}$H$_{6}$) as an example, the local Hamiltonian of a single carbon-carbon (C-C) bond in the benzene ring is determined by the local atomic environment, which includes the two carbon atoms involved in the bond, their attached hydrogen atoms, and the adjacent carbon atoms in the ring. The delocalized $\pi$-electron system, formed by the overlapping $p$-orbitals of the carbon atoms, creates a unique electronic environment that influences the local Hamiltonian. In contrast, atoms or bonds farther away from this local region, such as those in a substituent group attached to the benzene ring, have a much smaller impact on the local Hamiltonian of the C-C bond. Therefore, learning the local chemical environment is crucial for predicting the Hamiltonian.

To this end, we perform fine-grained cross-modal alignment, i.e., fragment-wise alignment, to build the correlations between string fragments and local geometries. We first segment SMILES strings into several fragments through BRICS~\cite{degen2008art}. According to the fragment, we segment the geometric embeddings $\mathbf{v}$ and token embeddings $\mathbf{t}^*$ into local atomic embeddings and $\{\mathbf{v}_{l_1},\ldots \mathbf{v}_{l_N}\}$ and local token embeddings $\{\mathbf{t}_{l_1}^*,\ldots \mathbf{t}_{l_N}^*\}$, where $N$ is the number of segmented fragments. Instead of a direct alignment for the two embeddings through a naive contrastive loss, we propose to contextualize the local token embeddings $\mathbf{t}_{l_i}^*$ with the local atomic embeddings $\mathbf{v}_{l_i}$ for a more contextualized string token-geometry alignment. The contextualized geometry-aware token embeddings $\mathbf{t}_{l}^{v}$ is produced by local token embeddings-conditioned attention pooling,
\begin{equation}	\label{cross_attention}
    \begin{aligned}
    \mathbf{t}_{l}^v = \texttt{softmax}\left( \frac{\mathbf{t}_l^* W_q ( \mathbf{v}_l^\top W_k)}{\sqrt{d}} \right) \mathbf{v}_l W_v,
    \end{aligned}
\end{equation}
where $W_q$, $W_k$, $W_v$ are the query, key, and value weight matrices, and $d$ is the dimension of the weight matrices.
We then adopt the contrastive loss to maximize the similarity between positive pairs $(\mathbf{v}_{l_i}, \mathbf{t}_{l_i}^v)$ and minimize the similarity between negative pairs $(\mathbf{v}_{l_i}, \mathbf{t}_{l_j}^v)$. We adopt a contrastive loss based on the sigmoid function as proposed by SigLIP \cite{zhai2023sigmoid}. This loss is favored over the conventional InfoNCE loss \cite{oord2018representation} primarily because it offers superior efficiency for multi-GPU training. Consequently, our local atomic environment-aware contrastive learning is defined as:
\begin{equation}
    \mathcal{L}^l_{i,j} = -\frac{1}{1+\exp({y_{i,j}(\cos(\mathbf{v}_{l_i}, \mathbf{t}_{l_j}^v)/\tau)) }},
\end{equation}
where $y_{i,j}$ is $+1$ for positive pairs when $i=j$ and $-1$ for negative pairs otherwise.
% $\mathcal{L}^g$ mainly optimizes the molecule-level embeddings, it is beneficial for global alignment. 
% We empirically find that combining $\mathcal{L}^g$ with $\mathcal{L}^l$ consistently improves performance. 
The final loss in the pretraining stage $\mathcal{L}$ is the sum of the fragment-level contrastive loss and the modality discrepancy loss, and it is defined by $\mathcal{L}=\mathcal{L}^D+\mathcal{L}^l$. This latent space alignment mechanism not only builds correlations between string fragments and local geometries but also forces the token embeddings to align with the geometric space, thereby facilitating the preservation of equivariance.

\subsection{Weakly Supervised Fine-Tuning Stage}

Following the pre-training stage, the fine-tuning process operates exclusively on SMILES strings. This stage employs a projection head to adapt the pre-trained SMILES-BERT for direct Hamiltonian prediction. However, the scarcity of Hamiltonian data could degrade the model's generalization capability. To address this limitation, we introduce a weakly supervised strategy. 
Weakly supervised learning mitigates the low-data issue by using training data with limited, imprecise, or incomplete label information~\cite{zhang2021weakly,shen2023survey}. In our work, we use a weakly supervised learning method leveraging a masking mechanism to fine-tune our model. Specifically, given a SMILES string $\mathbf{s}$, $N$ fragments $\mathbf{s} = \{s_1, ..., s_N\}$ can be identified by using BRICS~\cite{degen2008art}. We apply a randomly generated binary vector $\mathbf{m} \in \{0, 1\}^N$ to the fragments to mask the corresponding fragment $s_i$. We use the masked string $\tilde{\mathbf{s}}=\mathbf{s} \odot \mathbf{m}$ for predicting the entire Hamiltonian matrix. Both $\mathbf{s}$ and $\tilde{\mathbf{s}}$ will be taken to the SMILES-BERT, predicting their corresponding Hamiltonian matrix denoted as $\mathbf{H}$ and $\tilde{\mathbf{H}}$. MAE and MSE loss functions are adopted to minimize the prediction error,
% \begin{equation}
%         \mathcal{L} = \frac{1}{n} \sum_{i=1}^{n} |\mathbf{H}_i^* - \mathbf{H}_i| + \frac{1}{n} \sum_{i=1}^{n} (\mathbf{H}_i^* - \mathbf{H}_i)^2+ \frac{1}{n} \sum_{i=1}^{n} |\mathbf{H}_i^* - \tilde{\mathbf{H}}_i| + \frac{1}{n} \sum_{i=1}^{n} (\mathbf{H}_i^* - \tilde{\mathbf{H}}_i)^2,
% \end{equation}
\begin{equation}
\begin{split}
    \mathcal{L} = &\frac{\lambda_2}{n} \sum_{i=1}^{n} \left( |\mathbf{H}_i^* - \mathbf{H}_i| + (\mathbf{H}_i^* - \mathbf{H}_i)^2 \right)\\
    &+ \frac{1 - \lambda_2}{n} \sum_{i=1}^{n} \left( |\mathbf{H}_i^* - \tilde{\mathbf{H}}_i| + (\mathbf{H}_i^* - \tilde{\mathbf{H}}_i)^2 \right),
\end{split}
\end{equation}
% The masked parts  have a high probability of being replaced by $[MASK]$ or a low probability of being replaced by other fragments. After the operations, the masked parts and remained parts are recomposed as masked SMILES $\mathbf{s}^*$.
where $\mathbf{H}^*$ denotes the ground-truth Hamiltonian matrix and $\lambda_2$ is the hyperparameter that balances the diagonal and non-diagonal parts. Regarding the impact of the parameter $\lambda_2$ on the model's performance, please refer to Supplemental Information. 
The mask mechanism not only enhances model generalization under the low-data regime but also establishes a direct relationship between molecular fragments and their corresponding local Hamiltonians, enabling more accurate and generalized predictions. 

It should be noted that the Hamiltonian matrix is defined as a Hermitian matrix. In the real-space or real-coefficient basis sets, this property reduces to real symmetry. Therefore, to achieve efficiency, the matrix can be determined solely by calculating or predicting its upper triangular components and the diagonal, with the remaining elements being reconstructed via the symmetry operation.

%We leverage a calibrated Gaussian distribution to approximate the assumed Gaussian distribution, and thereby, there exists a distribution shift error, consisting of a distribution assumption error and a distribution approximation error. 
% As discussed before \cite{ben2010theory, blitzer2008learning}, setting $\tau$ introduces a trade-off between the support set that is reliable but not sufficient and the generated set that is sufficient but not reliable. Setting $\tau = 0.5$ means that we treat the generated set equally as the support set. 

% The predictor $\hat{f}$ is learned from $\mathcal{D}_{cd}$ and ,
% \begin{equation}
% \hat{f} = \arg \min_{f \in \mathcal{F}} \hat{R}_{(s+g)}(f)
% \end{equation}

%%%%%%%%%%%%%%%

\section{Experimental Studies}

\subsection{Experimental Setup}

 \textbf{Datasets}. We evaluate our model on four datasets: MD17~\cite{schutt2019unifying}, QH9~\cite{NEURIPS2023_7f755e27}, the newly curated QH-BM, and QH9-1000K. The MD17 dataset comprises four small molecules—water, ethanol, malondialdehyde, and uracil—each featuring thousands of 3D conformations. The QH9 dataset is organized into two subsets: QH9-stable, containing 130,831 small molecules, and QH9-dynamic, which includes 100 trajectory frames for each of 999 molecules. To assess model performance on larger systems, we constructed the QH-BM dataset, consisting of 5,446 molecules with more than 20 atoms each. Additionally, to test the model's robustness under high-temperature, non-equilibrium conditions, we assembled the QH9-1000K dataset, which contains molecular structures sampled at 1000K. These datasets encompass diverse experimental settings, including both in-distribution (id) and out-of-distribution (ood) splits, to rigorously validate the model's generalization capabilities. For further details on dataset collection and specifications, please refer to the Supplementary Material.

\begin{table*}[t]\scriptsize
  \caption{Overall performance comparison on the QH9 dataset. The unit for the Hamiltonian $\mathbf{H}$ and eigenenergies $\psi$ is Hartree denoted by $E_h$. The \colorbox{MyGreen4}{best} results and the \colorbox{MyGreen1}{second best} results are highlighted, respectively.}
  \label{tab:qh9_without_deqhnet}
  \centering
  \resizebox{0.95\linewidth}{!}{
    \begin{tabular}{ccccccc}
      \toprule
      \multirow{2}{*}{Dataset} & \multirow{2}{*}{Model} & \multicolumn{3}{c}{\(\mathbf{H}\) [$E_h$] $\downarrow$} & \multirow{2}{*}{$\epsilon$ [$E_h$] $\downarrow$} & \multirow{2}{*}{$\psi$ $\uparrow$} \\
      \cmidrule(lr){3-5}
      ~ & ~ & diagonal & non-diagonal & all & ~ & ~ \\
      \midrule
      \multirow{7}{*}{QH9-stable-id} & SMILES-BERT & $(2.08 \pm 0.27) \times 10^{-1}$ & $(5.63 \pm 0.50) \times 10^{-2}$ & $(6.83 \pm 0.77) \times 10^{-2}$ & $(8.03 \pm 0.21) \times 10^{1}$ & $(3.74 \pm 0.46) \times 10^{-2}$ \\
      ~ & QHNet & $(1.51 \pm 0.33) \times 10^{-1}$ & $(3.81 \pm 0.55) \times 10^{-2}$ & $(4.56 \pm 0.57) \times 10^{-2}$ & $(1.90 \pm 0.27) \times 10^{2}$ & $(3.80 \pm 0.48) \times 10^{-2}$ \\
      ~ & DEQHNet & \cellcolor{MyGreen1}$(1.08 \pm 0.08) \times 10^{-1}$ & \cellcolor{MyGreen1}$(3.77 \pm 0.02) \times 10^{-2}$ & \cellcolor{MyGreen1}$(4.31 \pm 0.07) \times 10^{-2}$ & $(2.40 \pm 0.51) \times 10^{2}$ & $(1.02 \pm 0.17) \times 10^{-1}$ \\
      ~ & SE(3)-Transformer & $(1.09 \pm 0.49) \times 10^{-1}$ & $(4.30 \pm 0.59) \times 10^{-2}$ & $(4.67 \pm 0.40) \times 10^{-2}$ & \cellcolor{MyGreen1}$(7.59 \pm 0.33) \times 10^{0}$ & \cellcolor{MyGreen4}$(1.32 \pm 0.56) \times 10^{-1}$ \\
      ~ & GemNet & $(3.60 \pm 0.44) \times 10^{-1}$ & $(4.37 \pm 0.62) \times 10^{-2}$ & $(6.13 \pm 0.33) \times 10^{-2}$ & \cellcolor{MyGreen4}$(6.61 \pm 0.18) \times 10^{0}$ & \cellcolor{MyGreen1}$(1.31 \pm 0.39) \times 10^{-1}$ \\
      ~ & {\em MGAHam} & \cellcolor{MyGreen4}$(5.15 \pm 1.19) \times 10^{-2}$ & \cellcolor{MyGreen4}$(3.76 \pm 0.01) \times 10^{-2}$ & \cellcolor{MyGreen4}$(3.84 \pm 0.05) \times 10^{-2}$ & $(6.74 \pm 0.43) \times 10^{1}$ & $(4.87 \pm 1.75) \times 10^{-2}$ \\
      \midrule
      \multirow{7}{*}{QH9-stable-ood} & SMILES-BERT & $(2.10 \pm 0.05) \times 10^{-1}$ & $(5.40 \pm 0.07) \times 10^{-2}$ & $(6.62 \pm 0.67) \times 10^{-2}$ & $(8.21 \pm 0.64) \times 10^{1}$ & $(3.96 \pm 0.20) \times 10^{-2}$ \\
      ~ & QHNet & $(2.58 \pm 1.19) \times 10^{-1}$ & $(3.50 \pm 0.32) \times 10^{-2}$ & $(4.75 \pm 0.52) \times 10^{-2}$ & $(3.05 \pm 0.86) \times 10^{2}$ & $(3.61 \pm 0.44) \times 10^{-2}$ \\
      ~ & DEQHNet & $(3.07 \pm 0.75) \times 10^{-1}$ & \cellcolor{MyGreen1}$(3.19 \pm 0.01) \times 10^{-2}$ & $(4.72 \pm 0.40) \times 10^{-2}$ & $(4.05 \pm 5.73) \times 10^{2}$ & \cellcolor{MyGreen4}$(1.74 \pm 0.44) \times 10^{-1}$ \\
      ~ & SE(3)-Transformer & \cellcolor{MyGreen1}$(9.55 \pm 0.39) \times 10^{-2}$ & $(3.57 \pm 0.37) \times 10^{-2}$ & \cellcolor{MyGreen1}$(3.82 \pm 0.56) \times 10^{-2}$ & \cellcolor{MyGreen1}$(6.45 \pm 0.47) \times 10^{0}$ & $(1.22 \pm 0.44) \times 10^{-1}$ \\
      ~ & GemNet & $(4.27 \pm 0.42) \times 10^{-1}$ & $(3.35 \pm 0.52) \times 10^{-2}$ & $(5.01 \pm 0.33) \times 10^{-2}$ & \cellcolor{MyGreen4}$(5.69 \pm 0.39) \times 10^{0}$ & \cellcolor{MyGreen1}$(1.16 \pm 0.25) \times 10^{-1}$ \\
      ~ & {\em MGAHam} & \cellcolor{MyGreen4}$(4.67 \pm 0.33) \times 10^{-2}$ & \cellcolor{MyGreen4}$(3.15 \pm 0.51) \times 10^{-2}$ & \cellcolor{MyGreen4}$(3.58 \pm 0.32) \times 10^{-2}$ & $(6.41 \pm 0.38) \times 10^{1}$ & $(4.95 \pm 1.77) \times 10^{-2}$ \\
      \midrule
      \multirow{7}{*}{QH9-dynamic-geo} & SMILES-BERT & $(2.14 \pm 0.09) \times 10^{-1}$ & $(5.57 \pm 0.53) \times 10^{-2}$ & $(6.83 \pm 0.79) \times 10^{-2}$ & $(8.04 \pm 0.31) \times 10^{1}$ & $(3.73 \pm 0.54) \times 10^{-2}$ \\
      ~ & QHNet & $(1.54 \pm 0.01) \times 10^{-1}$ & $(3.83 \pm 0.37) \times 10^{-2}$ & $(4.58 \pm 0.09) \times 10^{-2}$ & $(2.40 \pm 0.36) \times 10^{2}$ & $(3.77 \pm 0.35) \times 10^{-2}$ \\
      ~ & DEQHNet & $(2.18 \pm 0.44) \times 10^{-1}$ & \cellcolor{MyGreen4}$(3.75 \pm 0.74) \times 10^{-2}$ & $(4.98 \pm 0.66) \times 10^{-2}$ & $(1.84 \pm 2.60) \times 10^{2}$ & \cellcolor{MyGreen4}$(2.04 \pm 0.30) \times 10^{-1}$ \\
      ~ & SE(3)-Transformer & \cellcolor{MyGreen1}$(7.02 \pm 0.65) \times 10^{-2}$ & $(3.84 \pm 0.50) \times 10^{-2}$ & \cellcolor{MyGreen1}$(4.01 \pm 0.27) \times 10^{-2}$ & \cellcolor{MyGreen4}$(8.00 \pm 0.26) \times 10^{0}$ & $(1.34 \pm 0.40) \times 10^{-1}$ \\
      ~ & GemNet & $(3.68 \pm 0.56) \times 10^{-1}$ & $(4.02 \pm 0.67) \times 10^{-2}$ & $(5.75 \pm 0.55) \times 10^{-2}$ & $(7.96 \pm 0.53) \times 10^{1}$ & \cellcolor{MyGreen1}$(1.30 \pm 0.39) \times 10^{-1}$ \\
      ~ & {\em MGAHam} & \cellcolor{MyGreen4}$(5.25 \pm 0.13) \times 10^{-2}$ & \cellcolor{MyGreen1}$(3.81 \pm 0.08) \times 10^{-2}$ & \cellcolor{MyGreen4}$(3.96 \pm 0.07) \times 10^{-2}$ & \cellcolor{MyGreen1}$(6.70 \pm 0.09) \times 10^{1}$ & $(4.76 \pm 1.56) \times 10^{-2}$ \\
      \midrule
      \multirow{7}{*}{QH9-dynamic-mol} & SMILES-BERT & $(2.13 \pm 0.05) \times 10^{-1}$ & $(5.76 \pm 0.85) \times 10^{-2}$ & $(6.99 \pm 1.04) \times 10^{-2}$ & $(8.29 \pm 0.67) \times 10^{1}$ & $(3.71 \pm 0.22) \times 10^{-2}$ \\
      ~ & QHNet & \cellcolor{MyGreen1}$(1.47 \pm 0.01) \times 10^{-1}$ & \cellcolor{MyGreen1}$(3.87 \pm 0.44) \times 10^{-2}$ & $(4.59 \pm 0.52) \times 10^{-2}$ & $(2.41 \pm 0.27) \times 10^{2}$ & $(3.71 \pm 0.28) \times 10^{-2}$ \\
      ~ & DEQHNet & $(2.93 \pm 0.70) \times 10^{-1}$ & \cellcolor{MyGreen4}$(3.83 \pm 0.01) \times 10^{-2}$ & $(5.47 \pm 0.45) \times 10^{-2}$ & $(1.67 \pm 2.36) \times 10^{2}$ & \cellcolor{MyGreen4}$(1.87 \pm 0.43) \times 10^{-1}$ \\
      ~ & SE(3)-Transformer & $(2.56 \pm 0.65) \times 10^{-1}$ & $(2.55 \pm 0.58) \times 10^{-1}$ & \cellcolor{MyGreen4}$(2.56 \pm 0.28) \times 10^{-2}$ & \cellcolor{MyGreen4}$(3.50 \pm 0.39) \times 10^{0}$ & $(1.25 \pm 0.67) \times 10^{-1}$ \\
      ~ & GemNet & $(3.68 \pm 0.36) \times 10^{-1}$ & $(4.03 \pm 0.59) \times 10^{-2}$ & $(5.75 \pm 0.61) \times 10^{-2}$ & $(8.56 \pm 0.46) \times 10^{1}$ & \cellcolor{MyGreen1}$(1.32 \pm 0.36) \times 10^{-1}$ \\
      ~ & {\em MGAHam} & \cellcolor{MyGreen4}$(5.06 \pm 0.88) \times 10^{-2}$ & $(3.90 \pm 0.04) \times 10^{-2}$ & \cellcolor{MyGreen1}$(3.97 \pm 0.46) \times 10^{-2}$ & \cellcolor{MyGreen1}$(6.98 \pm 0.68) \times 10^{1}$ & $(4.95 \pm 1.57) \times 10^{-2}$ \\
      \bottomrule
    \end{tabular}
  }
\end{table*}

\begin{table*}[t]\scriptsize
  \caption{Overall performance comparison on the QH9-1000K and QH-BM datasets. The \colorbox{MyGreen4}{best} results and the \colorbox{MyGreen1}{second best} results are highlighted, respectively.}
  \label{tab:1000k}
  \centering
  \resizebox{0.95\linewidth}{!}{
    \begin{tabular}{ccccccc}
      \toprule
      \multirow{2}{*}{Dataset} & \multirow{2}{*}{Model} & \multicolumn{3}{c}{\(\mathbf{H}\) [$E_h$] $\downarrow$} & \multirow{2}{*}{$\epsilon$ [$E_h$] $\downarrow$} & \multirow{2}{*}{$\psi$ $\uparrow$} \\
      \cmidrule(lr){3-5}
      ~ & ~ & diagonal & non-diagonal & all & ~ & ~ \\
      \midrule
      \multirow{6}{*}{QH9-1000K-id} & SMILES-BERT & $(2.54 \pm 0.16) \times 10^{-1}$ & $(3.88 \pm 2.71) \times 10^{-2}$ & $(5.54 \pm 2.54) \times 10^{-2}$ & $(3.03 \pm 0.02) \times 10^{0}$ & $(5.36 \pm 0.86) \times 10^{-2}$ \\
      ~ & QHNet & \cellcolor{MyGreen1}$(1.79 \pm 0.40) \times 10^{-1}$ & $(5.75 \pm 0.02) \times 10^{-3}$ & \cellcolor{MyGreen1}$(1.91 \pm 0.06) \times 10^{-2}$ & \cellcolor{MyGreen1}$(1.58 \pm 0.16) \times 10^{0}$ & $(6.14 \pm 0.53) \times 10^{-2}$ \\
      ~ & DEQHNet & $(2.29 \pm 1.00) \times 10^{-1}$ & \cellcolor{MyGreen1}$(5.73 \pm 0.01) \times 10^{-3}$ & $(2.34 \pm 0.82) \times 10^{-2}$ & $(5.16 \pm 7.29) \times 10^{0}$ & \cellcolor{MyGreen1}$(1.49 \pm 0.89) \times 10^{-1}$ \\
      ~ & SE(3)-Transformer & $(2.12 \pm 0.21) \times 10^{-1}$ & $(1.55 \pm 0.15) \times 10^{-2}$ & $(3.07 \pm 0.31) \times 10^{-2}$ & $(2.91 \pm 0.29) \times 10^{0}$ & $(1.08 \pm 0.11) \times 10^{-1}$ \\
      ~ & GemNet & $(2.02 \pm 0.24) \times 10^{-1}$ & $(1.48 \pm 0.18) \times 10^{-2}$ & $(2.93 \pm 0.35) \times 10^{-2}$ & $(2.78 \pm 0.33) \times 10^{0}$ & $(1.03 \pm 0.12) \times 10^{-1}$ \\
      ~ & {\em MGAHam} & \cellcolor{MyGreen4}$(1.08 \pm 0.03) \times 10^{-1}$ & \cellcolor{MyGreen4}$(5.60 \pm 0.05) \times 10^{-3}$ & \cellcolor{MyGreen4}$(1.39 \pm 0.05) \times 10^{-2}$ & \cellcolor{MyGreen4}$(8.02 \pm 0.25) \times 10^{-1}$ & \cellcolor{MyGreen4}$(1.52 \pm 0.08) \times 10^{-1}$ \\
      \midrule
      \multirow{6}{*}{QH9-1000K-ood} & SMILES-BERT & $(2.33 \pm 0.50) \times 10^{-1}$ & $(3.57 \pm 2.64) \times 10^{-2}$ & $(4.98 \pm 2.96) \times 10^{-2}$ & \cellcolor{MyGreen1}$(3.56 \pm 0.86) \times 10^{0}$ & $(4.61 \pm 2.11) \times 10^{-2}$ \\
      ~ & QHNet & $(3.37 \pm 2.03) \times 10^{-1}$ & \cellcolor{MyGreen1}$(4.03 \pm 0.02) \times 10^{-3}$ & $(2.24 \pm 1.18) \times 10^{-2}$ & $(6.87 \pm 5.42) \times 10^{0}$ & $(4.99 \pm 1.18) \times 10^{-2}$ \\
      ~ & DEQHNet & \cellcolor{MyGreen1}$(1.94 \pm 1.19) \times 10^{-1}$ & \cellcolor{MyGreen4}$(3.98 \pm 0.05) \times 10^{-3}$ & \cellcolor{MyGreen1}$(1.55 \pm 0.73) \times 10^{-2}$ & $(9.35 \pm 13.22) \times 10^{0}$ & \cellcolor{MyGreen1}$(1.42 \pm 1.14) \times 10^{-1}$ \\
      ~ & SE(3)-Transformer & $(2.32 \pm 0.23) \times 10^{-1}$ & $(1.34 \pm 0.13) \times 10^{-2}$ & $(2.72 \pm 0.27) \times 10^{-2}$ & $(5.65 \pm 0.56) \times 10^{0}$ & $(9.54 \pm 0.95) \times 10^{-2}$ \\
      ~ & GemNet & $(2.21 \pm 0.27) \times 10^{-1}$ & $(1.28 \pm 0.15) \times 10^{-2}$ & $(2.59 \pm 0.31) \times 10^{-2}$ & $(5.39 \pm 0.65) \times 10^{0}$ & $(9.11 \pm 1.09) \times 10^{-2}$ \\
      ~ & {\em MGAHam} & \cellcolor{MyGreen4}$(7.92 \pm 4.15) \times 10^{-2}$ & $(4.92 \pm 1.73) \times 10^{-3}$ & \cellcolor{MyGreen4}$(1.11 \pm 0.41) \times 10^{-2}$ & \cellcolor{MyGreen4}$(7.51 \pm 1.00) \times 10^{-1}$ & \cellcolor{MyGreen4}$(1.45 \pm 0.27) \times 10^{-1}$ \\
      \midrule
      \multirow{6}{*}{QH-BM-id} & SMILES-BERT & $(3.01 \pm 0.10) \times 10^{-1}$ & $(3.08 \pm 0.15) \times 10^{-1}$ & $(3.00 \pm 0.25) \times 10^{-1}$ & $(7.73 \pm 0.19) \times 10^{1}$ & $(1.12 \pm 0.10) \times 10^{-1}$ \\
      ~ & QHNet & $(2.77 \pm 0.21) \times 10^{-1}$ & \cellcolor{MyGreen1}$(2.57 \pm 0.18) \times 10^{-1}$ & \cellcolor{MyGreen1}$(2.56 \pm 0.20) \times 10^{-1}$ & $(2.38 \pm 0.42) \times 10^{1}$ & $(1.52 \pm 0.08) \times 10^{-1}$ \\
      ~ & DEQHNet & \cellcolor{MyGreen1}$(2.38 \pm 0.18) \times 10^{-1}$ & $(2.72 \pm 0.19) \times 10^{-1}$ & $(2.68 \pm 0.21) \times 10^{-1}$ & \cellcolor{MyGreen1}$(1.42 \pm 0.25) \times 10^{1}$ & \cellcolor{MyGreen1}$(4.95 \pm 0.45) \times 10^{-1}$ \\
      ~ & SE(3)-Transformer & $(2.84 \pm 0.28) \times 10^{-1}$ & $(2.88 \pm 0.29) \times 10^{-1}$ & $(2.82 \pm 0.28) \times 10^{-1}$ & $(3.35 \pm 0.33) \times 10^{1}$ & $(4.63 \pm 0.46) \times 10^{-1}$ \\
      ~ & GemNet & $(2.71 \pm 0.32) \times 10^{-1}$ & $(2.75 \pm 0.33) \times 10^{-1}$ & $(2.70 \pm 0.32) \times 10^{-1}$ & $(3.20 \pm 0.38) \times 10^{1}$ & $(4.42 \pm 0.53) \times 10^{-1}$ \\
      ~ & {\em MGAHam} & \cellcolor{MyGreen4}$(2.15 \pm 0.40) \times 10^{-1}$ & \cellcolor{MyGreen4}$(2.10 \pm 0.07) \times 10^{-1}$ & \cellcolor{MyGreen4}$(2.03 \pm 0.16) \times 10^{-1}$ & \cellcolor{MyGreen4}$(2.05 \pm 0.88) \times 10^{1}$ & \cellcolor{MyGreen4}$(5.22 \pm 0.57) \times 10^{-1}$ \\
      \midrule
      \multirow{6}{*}{QH-BM-ood} & SMILES-BERT & $(3.26 \pm 0.95) \times 10^{-1}$ & $(3.28 \pm 0.96) \times 10^{-1}$ & $(3.16 \pm 0.92) \times 10^{-1}$ & $(1.01 \pm 0.30) \times 10^{2}$ & $(1.05 \pm 0.31) \times 10^{-1}$ \\
      ~ & QHNet & $(2.92 \pm 0.85) \times 10^{-1}$ & \cellcolor{MyGreen1}$(2.70 \pm 0.79) \times 10^{-1}$ & \cellcolor{MyGreen1}$(2.70 \pm 0.79) \times 10^{-1}$ & \cellcolor{MyGreen1}$(2.67 \pm 0.78) \times 10^{1}$ & $(1.46 \pm 0.43) \times 10^{-1}$ \\
      ~ & DEQHNet & \cellcolor{MyGreen1}$(2.51 \pm 0.73) \times 10^{-1}$ & $(2.85 \pm 0.83) \times 10^{-1}$ & $(2.83 \pm 0.83) \times 10^{-1}$ & $(2.73 \pm 0.47) \times 10^{1}$ & \cellcolor{MyGreen4}$(3.50 \pm 2.53) \times 10^{-1}$ \\
      ~ & SE(3)-Transformer & $(3.14 \pm 0.31) \times 10^{-1}$ & $(3.14 \pm 0.31) \times 10^{-1}$ & $(3.07 \pm 0.31) \times 10^{-1}$ & $(5.06 \pm 0.51) \times 10^{1}$ & \cellcolor{MyGreen1}$(3.44 \pm 0.34) \times 10^{-1}$ \\
      ~ & GemNet & $(2.99 \pm 0.36) \times 10^{-1}$ & $(3.00 \pm 0.36) \times 10^{-1}$ & $(2.93 \pm 0.35) \times 10^{-1}$ & $(4.83 \pm 0.58) \times 10^{1}$ & $(3.28 \pm 0.39) \times 10^{-1}$ \\
      ~ & {\em MGAHam} & \cellcolor{MyGreen4}$(2.48 \pm 0.79) \times 10^{-1}$ & \cellcolor{MyGreen4}$(2.59 \pm 0.76) \times 10^{-1}$ & \cellcolor{MyGreen4}$(2.47 \pm 0.72) \times 10^{-1}$ & \cellcolor{MyGreen4}$(2.63 \pm 1.65) \times 10^{1}$ & $(1.33 \pm 0.39) \times 10^{-1}$ \\
      \bottomrule
    \end{tabular}
  }
\end{table*}

\textbf{Baselines and Parameter Settings}. For the MD17 dataset, we compare {\em MGAHam} with SchNOrb \cite{schutt2019unifying}, PhiSNet \cite{unke2021se}, QHNet \cite{yu2023efficient}, DEQHNet \cite{wang2024infusing}, SE(3)-Transformer\cite{fuchs2020se}, and GemNet\cite{gasteiger2021gemnet}. For the QH9, QH9-1000K, and QH-BM datasets, due to the lack of orbital overlap, SchNOrb and PhiSNet cannot be applied to them. Therefore, we used QHNet, DEQHNet, SE(3)-Transformer, and GemNet for these datasets. Detailed parameter settings can be found in the Supplemental Material. 

\textbf{Evaluation Metrics}. We adopt three metrics to evaluate the model's performance. 
\begin{enumerate}
    \item MAE on Hamiltonian Matrix (\(\mathbf{H}\)): This measures the Mean Absolute Error (MAE) between the predicted Hamiltonian matrix and the ground-truth from DFT calculations, considering both diagonal and off-diagonal blocks that represent intra- and inter-atomic interactions, respectively. For the QH9 and QH-BM datasets, given the sparsity induced by distant atom pairs in larger molecules, we separately assess the MAE for the diagonal and off-diagonal blocks, as well as the total MAE of the matrix.
    \item MAE on Occupied Orbital Energies ($\epsilon$): The occupied orbital energies, including the HOMO and LUMO energies, reflect properties such as chemical reactivity, electronic excitation, and spectroscopic characteristics. These energies can be derived by diagonalizing the Hamiltonian matrix. We calculate their MAEs between the predicted and the ground-truth values.
    \item Cosine Similarity of Occupied Orbital Coefficients ($\psi$):  This evaluates the cosine similarity between the predicted and the ground-truth coefficients for occupied molecular orbitals to determine the similarity between predicted and reference electronic wavefunctions, which are crucial for inferring chemical properties. The coefficients for occupied molecular orbitals are also derived from the Hamiltonian matrix. To eliminate the physical ambiguity arising from the arbitrary phase of wavefunctions, we explicitly employ a phase-invariant metric: calculating the absolute value of the cosine similarity between the predicted and ground-truth orbitals, which ensures that model performance evaluation remains unaffected by sign flipping. Furthermore, for orbital alignment and degeneracy handling, we utilize the property that the $\texttt{torch.linalg.eigh}$ naturally outputs eigenvectors sorted by energy in ascending order, enabling us to adopt a direct alignment strategy based on energy ranking.
\end{enumerate}

In the experiments, we employed the $\texttt{PySCF}$ library to compute the basis function overlap matrix $S$ using the def2-SVP basis set. By integrating the Hamiltonian matrix predicted by our model, the generalized eigenvalue problem, i.e., Equation (\ref{eq:HCSC}) can be solved via Löwdin orthogonalization, where molecular coordinates are generated directly from SMILES via $\texttt{RDKit}$ without further force-field or quantum mechanical optimization. This workflow enables the rapid estimation of orbital energies and occupied orbital coefficients.

\subsection{Results and Discussion}

Given that acquiring both geometric and Hamiltonian data is computationally expensive, training data availability is often severely limited. The training-and-inference paradigm of MGAHam is specifically designed for such low-data regimes. To demonstrate this, we initially evaluated the performance of various methods under data-constrained settings. Specifically, for each dataset, models were trained on a randomly selected subset of 4,000 molecules and subsequently evaluated on standard test sets. All experiments were averaged over five independent trials to ensure statistical reliability.

\begin{figure*}[!ht]
	\centering
	\includegraphics[width=0.97\linewidth]{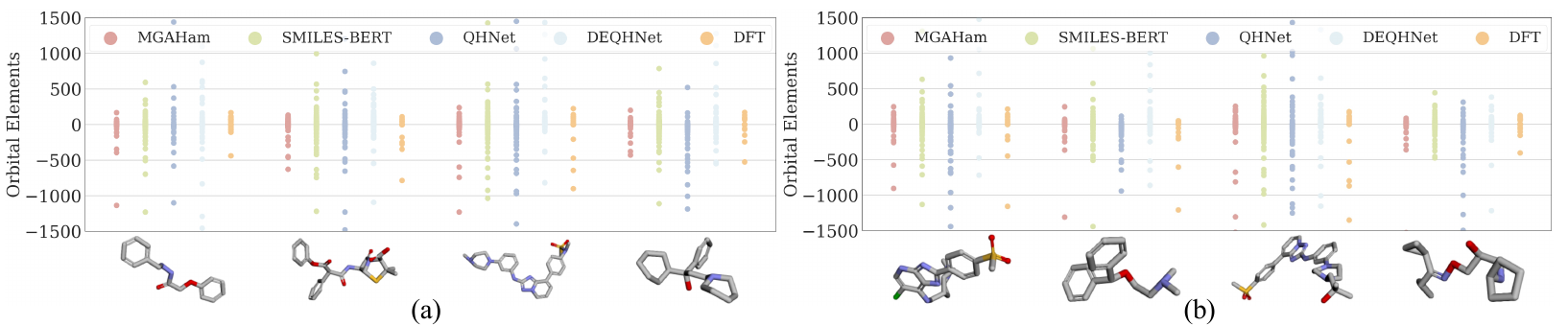}
  \caption{Orbital energy diagram of derived and DFT-calculated values for four large molecules.}
	\label{fig:orbital}
\end{figure*}

\begin{figure*}[!ht]
	\centering
	\includegraphics[width=1.0\linewidth]{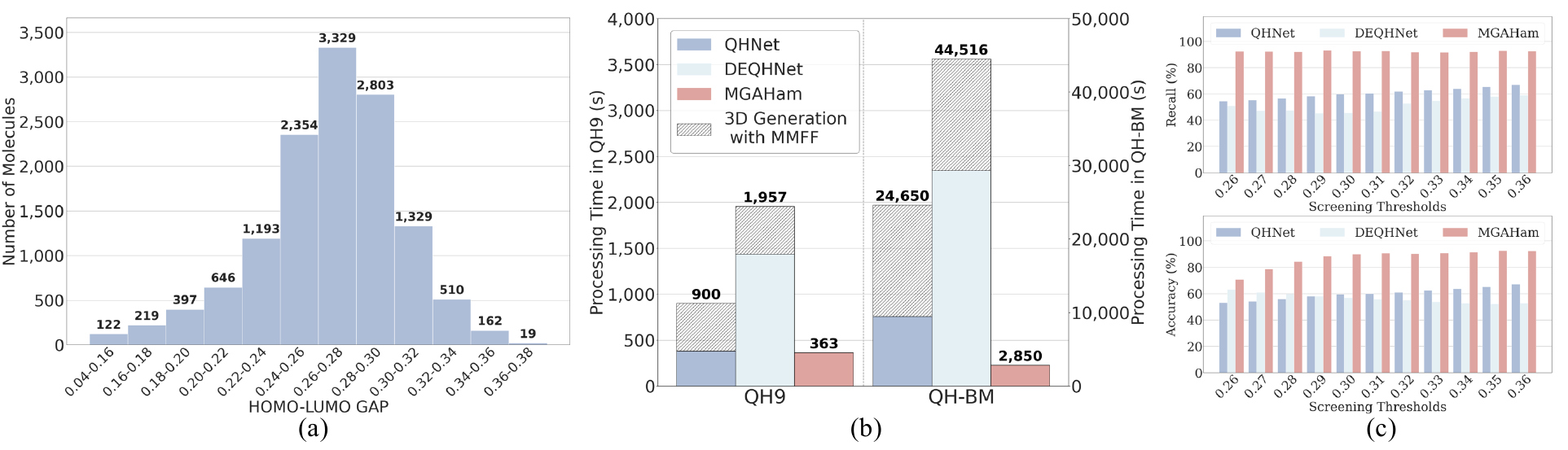}
 \caption{(a) Histogram of HOMO-LUMO gaps (0.02 eV bins) on the QH9 dataset. (b) The computation cost per 1,000 molecules on the QH9 and QH-BM datasets. The processing time includes the 3D structure generation cost using MMFF optimization and the model inference cost. (c) Prediction accuracy and recall across different screening thresholds.}
	\label{fig:vis2}
\end{figure*}

\textbf{Results on the QH9 dataset under the low-data regime.} Table \ref{tab:qh9_without_deqhnet} presents a comprehensive comparison of our proposed \textit{MGAHam} model against several baseline methods, including SMILES-BERT, QHNet, DEQHNet, SE(3)-Transformer, and GemNet across all settings of the QH9 dataset. 
A critical observation from the results is that \textit{MGAHam} consistently achieves comparable, and often superior, performance compared to other baselines across most test cases. For instance, on the QH9-stable-id (in-distribution setting), \textit{MGAHam} often exhibits the lowest MAE for Hamiltonian matrix elements (highlighted in red), significantly outperforming both the purely SMILES-based SMILES-BERT and the purely geometry-based methods, including QHNet, DEQHNet, SE(3)-Transformer, and GemNet. 
Even on the out-of-distribution setting, i.e., QH9-stable-ood dataset, \textit{MGAHam} obtains the lowest MAE for diagonal $\mathbf{H}$, non-diagonal $\mathbf{H}$, and all elements $\mathbf{H}$. This indicates its superior generalization in predicting Hamiltonian matrices. 

It is worth noting that, unlike QH9-stable-id and QH9-stable-ood, each molecule in QH9-dynamic-geo and QH9-dynamic-mol has one hundred conformations. These conformational variations lead to subtle differences in the Hamiltonian.
Consequently, when predicting Hamiltonians using only SMILES strings—which lack specific conformational details—the accuracy is inherently lower than that achieved by GNN counterparts that utilize 3D coordinates. We address potential strategies to alleviate this limitation in the subsequent Discussion on Scalability section.

\textbf{Results on the QH9-1000K and QH-BM datasets under the low-data regime.} The scarcity of Hamiltonian data and corresponding molecular geometries often poses a significant challenge in specialized research domains, particularly when characterizing electronic structures under extreme conditions—such as high temperatures—or when analyzing large, complex molecular systems. To rigorously evaluate our model's performance and generalization capabilities under such data-constrained conditions, we conducted assessments on the QH9-1000K and QH-BM datasets.

As can be seen from Table \ref{tab:1000k}, {\em MGAHam} achieved the best performance across almost all key metrics on all three datasets (QH9-1000K-id, QH9-1000K-ood, QH-BM-id, and QH-BM-ood). Here, we utilize two evaluation protocols on the $\text{QH-1000K}$ dataset: a random split and a size-ood split for evaluating generalization. The size-ood approach segregates the data by atom count, training the model only on molecules where $\text{N} < 20$ and testing on molecules where $\text{N} > 23$. The $\text{QH-BM-id}$ employs a random, in-distribution split for training and testing. For cross-element generalization assessment, $\text{QH-BM-ood}$ implements an element-wise separation: the training set is composed of molecules excluding Sulfur ($\text{S}$) and Phosphorus ($\text{P}$), while the testing set is built exclusively from molecules that contain these two specific elements.

On the QH9-1000K-id dataset, {\em MGAHam} demonstrated the best performance in Hamiltonian errors. {\em MGAHam}'s energy prediction error was only $0.802 E_h$ and achieved the highest wavefunction similarity $0.12$. 
On the QH9-1000K-ood dataset, {\em MGAHam}'s diagonal error (7.92 $\times 10^{-2}E_h$) is much lower than the next best DEQHNet, and its all error is as low as 1.11 $\times 10^{-2}E_h$. Its energy prediction error is only $0.751 E_h$, and wavefunction similarity reaches $1.45 \times 10^{-1}$, again surpassing all baselines. On the large molecular QH-BM dataset, {\em MGAHam} maintained its promising performance. {\em MGAHam} almost performed best in Hamiltonian errors, energy prediction error, and wavefunction similarity. 
These results demonstrate that {\em MGAHam} exhibits outstanding performance for handling molecular behaviors under specific scenarios, where Hamiltonian data and molecular geometries are scarce.

\begin{table*}[t]\scriptsize
  \caption{Overall performance comparison on the QH9 datasets, in which all models are trained using the standard data protocol established in the QH9 benchmark.}
  \label{tab:qh9_rich}
  \centering
  \resizebox{0.8\linewidth}{!}{
  \begin{tabular}{ccccccc}
    \toprule
    % --- 这是修改后的表头 ---
    \multirow{2}{*}{Dataset} & \multirow{2}{*}{Model} & \multicolumn{3}{c}{$H$[$10^{-6} E_h$] $\downarrow$} & \multirow{2}{*}{$\epsilon$[$10^{-6} E_h$] $\downarrow$} & \multirow{2}{*}{$\psi$[$10^{-2}$] $\uparrow$} \\
    \cmidrule(lr){3-5} 
    ~ & ~ & diagonal & non-diagonal & all & ~ & ~ \\
    % --- 修改结束 ---
\midrule
        \multirow{5}{*}{QH9-stable-id} & QHNet & $111.42 \pm 3.15$ & $73.78 \pm 2.04$ & $76.35 \pm 1.92$ & $799.21 \pm 22.40$ & $95.9 \pm 1.25$ \\
        ~ & DEQHNet & $108.91 \pm 2.84$ & $71.25 \pm 1.78$ & $73.44 \pm 1.62$ & $775.32 \pm 19.85$ & $94.8 \pm 1.13$ \\
        ~ & SE(3)-Transformer & $124.56 \pm 5.12$ & $82.15 \pm 2.45$ & $85.02 \pm 2.38$ & $985.44 \pm 35.20$ & $91.2 \pm 1.05$ \\
        ~ & GemNet & $118.92 \pm 4.58$ & $78.44 \pm 2.18$ & $81.12 \pm 2.24$ & $982.67 \pm 28.15$ & $93.1 \pm 1.18$ \\
        ~ & {\em MGAHam} & $113.67 \pm 4.22$ & $73.04 \pm 1.98$ & $75.72 \pm 2.15$ & $953.56 \pm 30.12$ & $94.8 \pm 1.13$ \\
        \midrule
        \multirow{5}{*}{QH9-stable-ood} & QHNet & $112.19 \pm 3.38$ & $69.94 \pm 2.11$ & $72.18 \pm 1.88$ & $644.87 \pm 18.75$ & $93.7 \pm 1.31$ \\
        ~ & DEQHNet & $109.54 \pm 2.95$ & $67.42 \pm 1.82$ & $69.85 \pm 1.54$ & $625.18 \pm 15.22$ & $92.1 \pm 1.19$ \\
        ~ & SE(3)-Transformer & $128.12 \pm 5.42$ & $76.88 \pm 2.54$ & $79.56 \pm 2.62$ & $997.73 \pm 32.11$ & $88.5 \pm 1.12$ \\
        ~ & GemNet & $122.45 \pm 4.88$ & $74.22 \pm 2.30$ & $76.84 \pm 2.44$ & $995.12 \pm 29.56$ & $90.4 \pm 1.08$ \\
        ~ & {\em MGAHam} & $118.54 \pm 4.95$ & $72.31 \pm 2.25$ & $74.79 \pm 2.41$ & $970.32 \pm 27.80$ & $92.5 \pm 1.09$ \\
        \bottomrule
  \end{tabular}}
\end{table*}

\textbf{Results on the QH9 dataset under the standard data regime.} In addition to the low-data experiments, we conducted a performance comparison adhering to the standard data protocol established in the QH9 benchmark\cite{NEURIPS2023_7f755e27}. The results are shown in Table \ref{tab:qh9_rich}. 
The performance comparison on the QH9 benchmark demonstrates that our proposed method, {\em MGAHam}, achieves competitive precision compared to state-of-the-art geometry-based models. Notably, while models like QHNet and DEQHNet rely on explicit 3D geometric descriptors, {\em MGAHam} bypasses complex 3D geometric calculations yet maintains a comparable level of accuracy. In the QH9-stable-id task, {\em MGAHam} yields an overall Hamiltonian error ($H_{all}$) of $75.72 \pm 2.15$. Furthermore, {\em MGAHam} outperforms established 3D geometric baselines such as SE(3)-Transformer ($85.02$) and GemNet ($81.12$). A similar trend is observed in the ood settings, where {\em MGAHam} demonstrates robust generalization with a $\psi$ score of $92.5$, closely matching the geometric counterparts. These results suggest that by effectively leveraging the molecular language, {\em MGAHam} can circumvent the computational overhead associated with 3D geometries without substantial degradation in predictive performance.

% On the other hand, the results of both QHNet and {\em MGAHam} clearly show a significant degradation of the QH9-1000K and QH-BM datasets. This suggests that while both models struggle with the complexity of these high-temperature, non-equilibrium, and larger molecular systems. Therefore, future efforts should focus on building more generalizable models to handle these complex systems.

 % The comparison in Table \ref{tab:qh9_rich} shows that QHNet generally delivers better accuracy, especially on QH9-stable. However, when testing on the more challenging high-temperature ($1000\text{K}$) and QH-BM datasets, the advantage of QHNet narrows. While the models show comparable Hamiltonian errors on the $1000\text{K}$ set, {\em MGAHam} unexpectedly shows lower diagonal Hamiltonian error on the OOD $1000\text{K}$ data. QHNet and {\em MGAHam} maintains a similar occupied orbital energies and cosine similarity of occupied orbital coefficients. This suggests that while both models struggle with the complexity of these high-temperature, non-equilibrium systems.

% This minor performance gap is understandable, as the BERT-based architecture using SMILES inherently struggles to represent and learn subtle geometric relationships. With a sufficient amount of data, GNNs can better capture these crucial geometric features. 
% However, we believe the combination of comparable predictive accuracy and extremely high predictive efficiency of our method is of great value. In the following sections, we will leverage this characteristic for high-throughput screening.

\subsection{High-Throughput Molecular Screening}

One notable feature of {\em MGAHam} is its ability to predict Hamiltonians using only SMILES strings, without requiring expensive molecular geometries. This feature makes it particularly well-suited for high-throughput screening of Hamiltonian-related properties. We will investigate this capability of {\em MGAHam} in this section.

\textbf{Screening Accuracy.} For high-throughput screening, the HOMO-LUMO energy gap is designated as the screening property. This electronic property reflects many macroscopic molecular behaviors. For instance, a low HOMO-LUMO gap indicates a molecule's increased propensity to donate electrons from its HOMO or accept electrons into its LUMO during chemical processes, i.e., a high chemical reactivity. Informed by the distribution of the HOMO-LUMO gap within the QH9 dataset (Fig. \ref{fig:vis2} (a)), we formally conceptualize the screening task as a series of binary classification problems by setting a range of screening thresholds, spanning from 0.26 eV to 0.36 eV. Specifically, the molecules exhibiting a HOMO-LUMO gap exceeding a defined threshold are categorized as the positive class, with all others assigned to the negative class. This range ensures comprehensive coverage of the gap distribution across both balanced and imbalanced classes, permitting a comprehensive assessment of the model's screening performance.

As illustrated in Fig. \ref{fig:vis2} (c), \textit{MGAHam} consistently demonstrates superior performance in terms of accuracy and recall when compared to both QHNet and DEQHNet across all screening thresholds. Notably, \textit{MGAHam} sustains a stable predictive capability (accuracy $> 0.70$, recall $> 0.90$) across the thresholds. This proven screening capability makes it an effective tool for accurate molecular Hamiltonian inference and accelerating virtual screening tasks.

\begin{figure*}[!ht]
	\centering
	\includegraphics[width=0.8\linewidth]{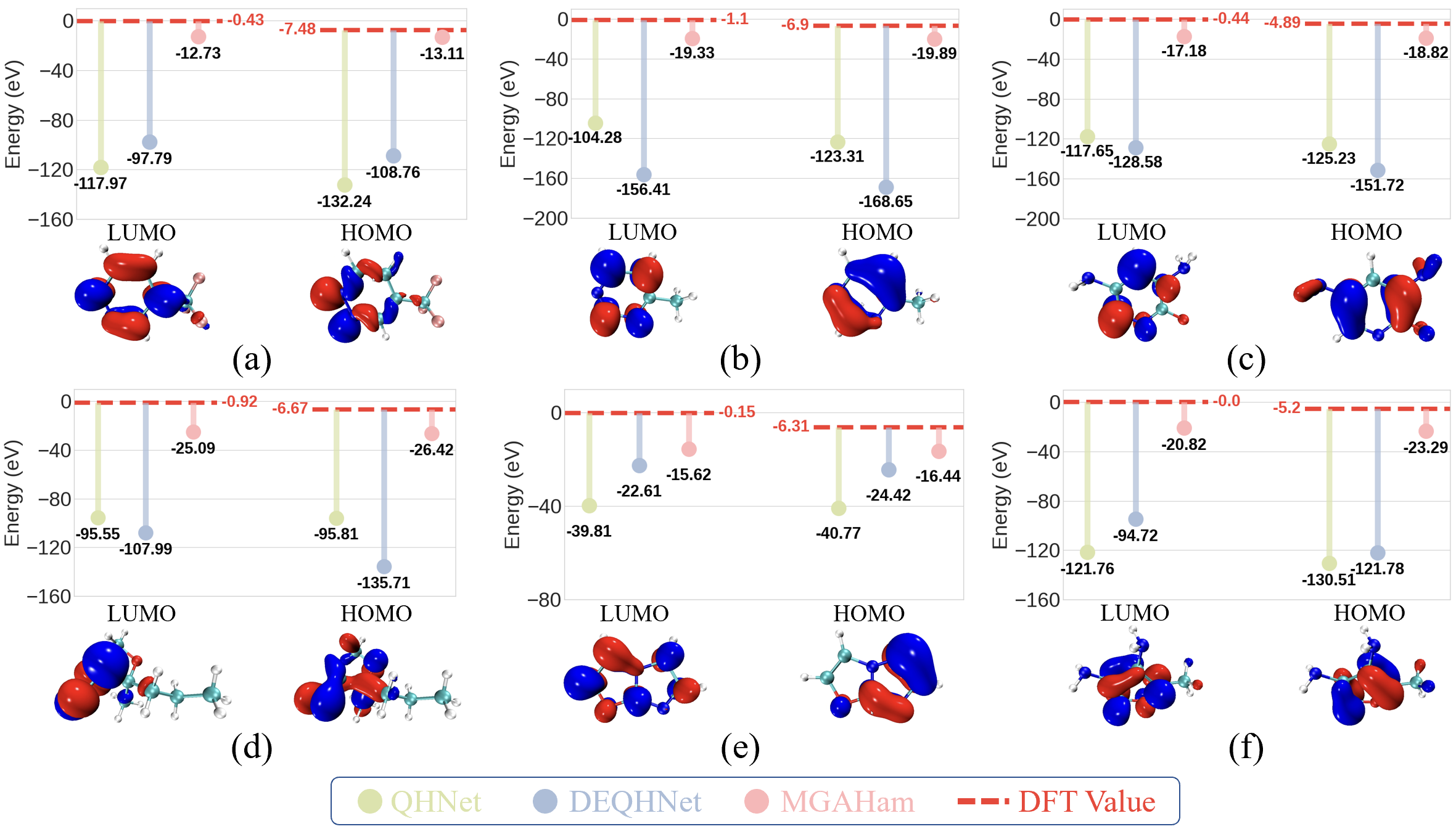}
	\caption{The derived and DFT-calculated HOMO and LUMO values of six molecules from the QH-BM dataset.}
	\label{fig:HOMO&LUMO}
\end{figure*}

We further visualized molecular orbital energies in Fig. \ref{fig:orbital}. These figures directly compare the orbital energy elements derived from the Hamiltonian predicted by \textit{MGAHam}, SMILES-BERT, QHNet, DEQHNet, and benchmark DFT calculations. These visualizations confirm that \textit{MGAHam}'s predicted orbital energies align significantly better with the DFT calculations than those from other models. To provide deeper insight into these critical electronic properties, we also visualize them. Fig. \ref{fig:HOMO&LUMO} showcases the derived HOMO and LUMO values for six representative molecules from the QH-BM dataset. These visual results show that \textit{MGAHam}'s predictions align closely with the corresponding DFT calculations.

\begin{figure}[t]
    \centering
    \includegraphics[width=0.98\linewidth]{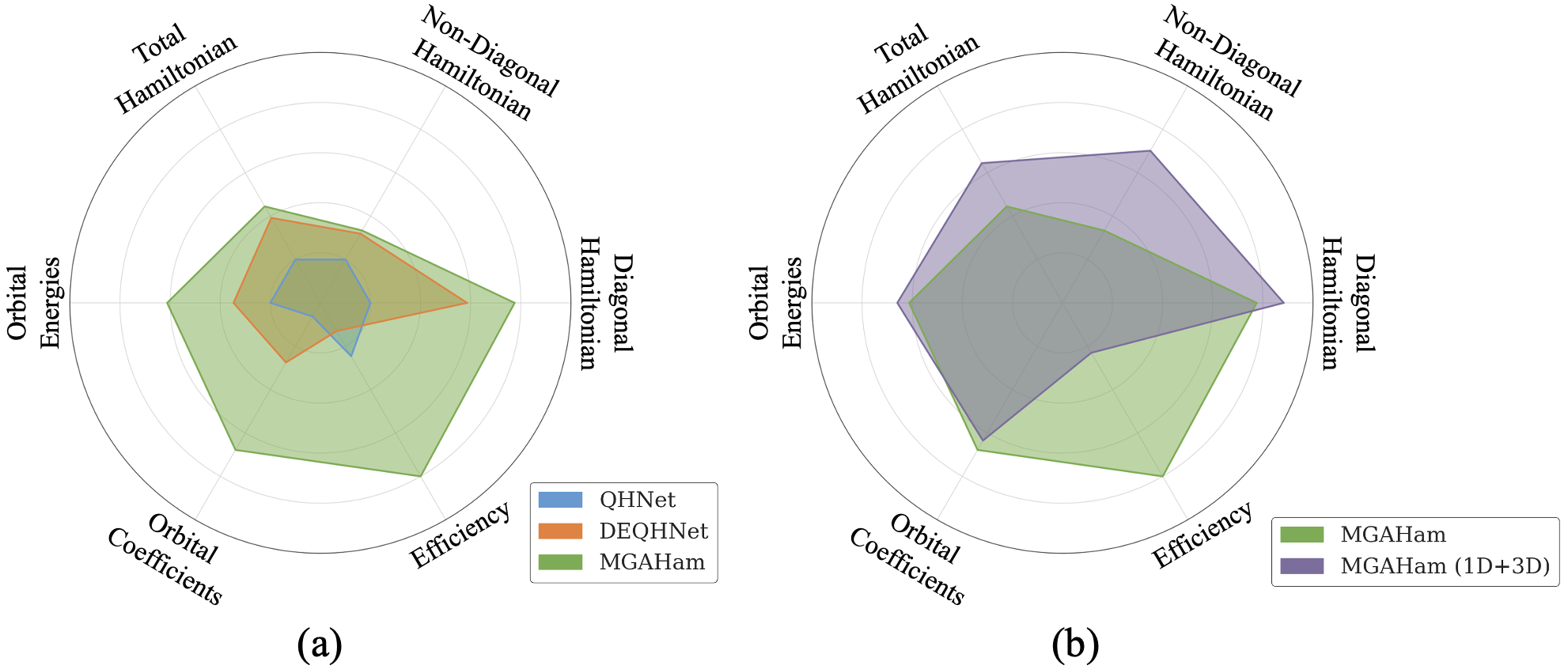}
    \caption{Comparative performance of our proposed {\em MGAHam}. (a) {\em MGAHam} surpasses QHNet and DEQHNet in terms of Hamiltonian, orbital energy, occupied orbital coefficients, and computational efficiency. (b). Compared with {\em MGAHam}, {\em MGAHam} (1D+3D) sacrifices computational efficiency, but it achieves higher accuracy in quantum property prediction.}
    \label{fig:radar}
\end{figure}

\begin{table}[!t]
   \caption{The acceleration ratio of QHNet, DEQHNet, and {\em MGAHam} compared to DFT, per 1,000 molecules on the QH9 and QH-BM datasets.}
    \centering
     \resizebox{0.97\linewidth}{!}{
    \begin{tabular}{c c l r}
    \toprule
        Dataset & DFT Calculation Time & Model & \multicolumn{1}{c}{Acceleration Ratio} \\
        \midrule
        \multirow{3}{*}{QH9} & \multirow{3}{*}{11.6 hours} & QHNet & 35.51$\times$ \\
        ~ & ~ & DEQHNet & 18.73$\times$ \\
        ~ & ~ & {\em MGAHam} & 115.42$\times$ \\
        \midrule
        \multirow{3}{*}{QH-BM} & \multirow{3}{*}{87.1 hours} & QHNet & 30.58$\times$ \\
        ~ & ~ & DEQHNet & 10.69$\times$ \\
        ~ & ~ & {\em MGAHam} & 110.07$\times$ \\
        \bottomrule
    \end{tabular}
    }
    \vspace{-0.1cm}
    \label{tab:accel}
\end{table}

\begin{table}[!t]
  \centering
  \caption{The computation cost of QHNet, DEQHNet, and {\em MGAHam} per 1,000 molecules on the QH9 and QH-BM datasets. The unit is second. }
  \resizebox{0.97\linewidth}{!}{
    \begin{tabular}{ccccc}
    \toprule
    Datasets & Time Cost & QHNet & DEQHNet & {\em MGAHam} \\
    \midrule
    \multirow{3}[4]{*}{QH9} & MMFF Time & 520   & 520   &  --- \\
          & Inference Time & 480   & 1,437 & 363 \\
\cmidrule{2-5}          & Total Time & 900   & 1,957 & 363 \\
    \midrule
    \multirow{3}[4]{*}{QH-BM} & MMFF Time & 15,190 & 15,190 & --- \\
          & Inference Time & 9,460 & 29,326 & 2,850 \\
\cmidrule{2-5}          & Total Time & 24,650 & 44,516 & 2,850 \\
    \bottomrule
    \end{tabular}}%
     \vspace{-0.2cm}
  \label{tab:costtime}%
\end{table}%

\textbf{Screening Efficiency.} Computational efficiency is a crucial factor in high-throughput virtual screening. Fig. \ref{fig:vis2} (b) clearly illustrates the computational time costs of the compared algorithms. Table \ref{tab:costtime} provides a detailed display of the various time components involved in the time cost. It is important to note that for baseline methods such as QHNet and DEQHNet, the reported times include the necessary overhead of acquiring molecular geometries. This typically involves computationally intensive steps like molecular dynamics calculations, e.g., here using the Merck molecular force field (MMFF), which contribute to their longer overall runtime. In contrast, \textit{MGAHam} completely bypasses this computationally demanding geometry acquisition step. This fundamental architectural advantage makes \textit{MGAHam} substantially faster than other baselines for virtual screening, particularly in practical scenarios where a molecule's geometry is not immediately available.

\begin{figure}[!t]
	\centering
	\includegraphics[width=1.0\linewidth]{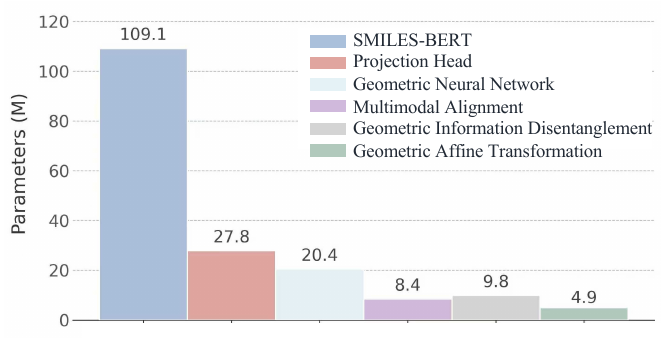}
 \caption{Parameter size of individual modules within \textit{MGAHam}.}
	\label{fig:mgaPara}
     \vspace{-0.4cm}
\end{figure}

\begin{table*}[t]
    \centering
    \caption{Overall performance comparison on the MD17 dataset. The unit for Hamiltonian \(\mathbf{H}\) and eigen energies $\epsilon$ is Hartree, denoted by $E_h$. The \colorbox{MyGreen4}{best} results and the \colorbox{MyGreen1}{second best} results are highlighted, respectively.}
    \label{tab:md17}
    \resizebox{0.76\textwidth}{!}{
        \begin{tabular}{ccccc}
            \toprule
            Dataset & Model & \(\mathbf{H}\) [$E_h$] $\downarrow$ & $\epsilon$ [$E_h$] $\downarrow$ & $\psi$ $\uparrow$\\
            \midrule
            \multirow{7}{*}{Water} & SchNOrb & $(4.26 \pm 0.15) \times 10^{-2}$ & $(3.65 \pm 0.50) \times 10^{-1}$ & $(3.84 \pm 0.13) \times 10^{-1}$ \\
            ~ & PhiSNet & $(2.24 \pm 0.20) \times 10^{-1}$ & $(2.81 \pm 1.34) \times 10^{0}$ & $(2.19 \pm 1.15) \times 10^{-1}$ \\
            ~ & QHNet & $(2.23 \pm 0.40) \times 10^{-1}$ & $(1.03 \pm 0.10) \times 10^{1}$ & $(6.30 \pm 0.10) \times 10^{-2}$ \\
            ~ & DEQHNet & $(1.27 \pm 0.40) \times 10^{-1}$ & $(1.17 \pm 0.20) \times 10^{0}$ & $(2.75 \pm 0.20) \times 10^{-1}$ \\
            ~ & SE(3)-Transformer & $(2.73 \pm 0.75) \times 10^{-1}$ & \cellcolor{MyGreen1}$(1.61 \pm 0.44) \times 10^{-1}$ & \cellcolor{MyGreen1}$(5.75 \pm 1.57) \times 10^{-1}$ \\
            ~ & GemNet & \cellcolor{MyGreen1}$(1.14 \pm 0.31) \times 10^{-1}$ & $(4.25 \pm 1.16) \times 10^{-1}$ & $(4.08 \pm 1.12) \times 10^{-1}$ \\
            ~ & {\em MGAHam} (1D+3D) & \cellcolor{MyGreen4}$(1.73 \pm 0.18) \times 10^{-2}$ & \cellcolor{MyGreen4}$(1.42 \pm 0.17) \times 10^{-1}$ & \cellcolor{MyGreen4}$(6.68 \pm 0.15) \times 10^{-1}$ \\
            \midrule
            \multirow{7}{*}{Ethanol} & SchNOrb & $(2.11 \pm 0.08) \times 10^{-2}$ & $(1.04 \pm 0.74) \times 10^{0}$ & \cellcolor{MyGreen4}$(2.78 \pm 1.26) \times 10^{-1}$ \\
            ~ & PhiSNet & $(8.50 \pm 0.02) \times 10^{-2}$ & $(4.19 \pm 3.66) \times 10^{0}$ & $(1.31 \pm 0.29) \times 10^{-1}$ \\
            ~ & QHNet & $(4.96 \pm 1.07) \times 10^{-2}$ & $(4.79 \pm 4.23) \times 10^{0}$ & $(1.35 \pm 0.96) \times 10^{-1}$ \\
            ~ & DEQHNet & $(4.35 \pm 0.74) \times 10^{-2}$ & $(1.44 \pm 1.56) \times 10^{1}$ & $(6.00 \pm 2.26) \times 10^{-2}$ \\
            ~ & SE(3)-Transformer & \cellcolor{MyGreen1}$(1.80 \pm 0.49) \times 10^{-2}$ & \cellcolor{MyGreen4}$(2.15 \pm 0.59) \times 10^{-1}$ & \cellcolor{MyGreen1}$(2.65 \pm 0.72) \times 10^{-1}$ \\
            ~ & GemNet & $(1.61 \pm 0.44) \times 10^{-1}$ & $(1.82 \pm 0.50) \times 10^{0}$ & $(2.04 \pm 0.56) \times 10^{-1}$ \\
            ~ & {\em MGAHam} (1D+3D) & \cellcolor{MyGreen4}$(1.36 \pm 0.18) \times 10^{-2}$ & \cellcolor{MyGreen1}$(4.72 \pm 0.73) \times 10^{-1}$ & $(2.18 \pm 0.03) \times 10^{-1}$ \\
            \midrule
            \multirow{7}{*}{Malonaldehyde} & SchNOrb & $(2.56 \pm 0.06) \times 10^{-2}$ & $(2.16 \pm 1.23) \times 10^{0}$ & $(1.77 \pm 0.65) \times 10^{-1}$ \\
            ~ & PhiSNet & $(8.24 \pm 0.02) \times 10^{-2}$ & $(6.67 \pm 6.97) \times 10^{0}$ & $(9.44 \pm 0.41) \times 10^{-2}$ \\
            ~ & QHNet & $(5.49 \pm 0.40) \times 10^{-2}$ & $(3.90 \pm 0.20) \times 10^{0}$ & $(1.07 \pm 0.10) \times 10^{-1}$ \\
            ~ & DEQHNet & $(5.05 \pm 0.40) \times 10^{-2}$ & $(1.35 \pm 0.03) \times 10^{1}$ & $(4.52 \pm 0.19) \times 10^{-2}$ \\
            ~ & SE(3)-Transformer & $(1.80 \pm 0.49) \times 10^{-2}$ & $(2.00 \pm 0.55) \times 10^{0}$ & \cellcolor{MyGreen4}$(2.00 \pm 0.55) \times 10^{-1}$ \\
            ~ & GemNet & \cellcolor{MyGreen1}$(1.60 \pm 0.44) \times 10^{-2}$ & \cellcolor{MyGreen1}$(1.80 \pm 0.49) \times 10^{0}$ & $(1.90 \pm 0.52) \times 10^{-1}$ \\
            ~ & {\em MGAHam} (1D+3D) & \cellcolor{MyGreen4}$(1.36 \pm 0.43) \times 10^{-2}$ & \cellcolor{MyGreen4}$(9.35 \pm 0.26) \times 10^{-1}$ & \cellcolor{MyGreen1}$(1.95 \pm 0.10) \times 10^{-1}$ \\
            \midrule
            \multirow{7}{*}{Uracil} & SchNOrb & \cellcolor{MyGreen1}$(1.17 \pm 0.17) \times 10^{-2}$ & $(1.81 \pm 2.00) \times 10^{0}$ & $(1.15 \pm 0.37) \times 10^{-1}$ \\
            ~ & PhiSNet & $(6.18 \pm 0.08) \times 10^{-2}$ & $(7.33 \pm 7.87) \times 10^{0}$ & $(8.90 \pm 1.84) \times 10^{-2}$ \\
            ~ & QHNet & $(5.72 \pm 0.40) \times 10^{-2}$ & $(1.80 \pm 0.40) \times 10^{0}$ & \cellcolor{MyGreen1}$(2.03 \pm 0.20) \times 10^{-1}$ \\
            ~ & DEQHNet & $(3.05 \pm 2.57) \times 10^{-2}$ & $(4.24 \pm 1.51) \times 10^{0}$ & $(6.02 \pm 1.95) \times 10^{-2}$ \\
            ~ & SE(3)-Transformer & $(1.36 \pm 0.37) \times 10^{-2}$ & \cellcolor{MyGreen1}$(1.79 \pm 0.36) \times 10^{0}$ & \cellcolor{MyGreen4}$(2.90 \pm 0.79) \times 10^{-1}$ \\
            ~ & GemNet & $(1.03 \pm 0.28) \times 10^{-1}$ & $(2.42 \pm 0.66) \times 10^{0}$ & $(2.09 \pm 0.57) \times 10^{-1}$ \\
            ~ & {\em MGAHam} (1D+3D) & \cellcolor{MyGreen4}$(1.12 \pm 0.21) \times 10^{-2}$ & \cellcolor{MyGreen4}$(1.28 \pm 0.71) \times 10^{0}$ & $(1.43 \pm 0.08) \times 10^{-1}$ \\
            \bottomrule
        \end{tabular}
    }
    \vspace{-0.1cm}
\end{table*}

Further quantifying this efficiency, Table \ref{tab:accel} presents the acceleration rates of the models compared to direct DFT calculations. For instance, calculating the Hamiltonian using DFT for just 1,000 small molecules on the QH9 dataset demands approximately 11.6 hours. This computational burden escalates dramatically for larger molecules, with 1,000 molecules on the QH-BM dataset requiring over 87 hours.
\textit{MGAHam}, on the other hand, offers a significant reduction in computational time. It completes the Hamiltonian prediction for 1,000 molecules on the QH9 dataset in only about 6 minutes ($\sim$0.1 hours) and for the more complex QH-BM dataset in roughly 0.8 hours. Comparing the acceleration ratios from Table \ref{tab:accel}, \textit{MGAHam} achieves the most significant speedup across all evaluated models. For the QH9 dataset, \textit{MGAHam} is approximately $\sim$115.42$\times$ faster than DFT, a stark contrast to QHNet's 35.51$\times$ and DEQHNet's 18.73$\times$. Similarly, on the QH-BM dataset, \textit{MGAHam} maintains an impressive acceleration of $\sim$110.07$\times$ compared to DFT, while QHNet achieves 30.58$\times$ and DEQHNet trails at 10.69$\times$. These figures demonstrate that \textit{MGAHam} operates 3 to 5 times faster than QHNet and DEQHNet.

\begin{table*}[!ht]
  \caption{Overall performance comparison on the QH9 and QH-BM dataset. The unit for the Hamiltonian $\mathbf{H}$ and eigen energies $\psi$ is Hartree denoted by $E_h$.}
  \label{tab:1dvs3d}
  \centering
  \resizebox{0.95\linewidth}{!}{
    \begin{tabular}{ccccccc}
      \toprule
      \multirow{2}{*}{Dataset} & \multirow{2}{*}{Model} & \multicolumn{3}{c}{\(\mathbf{H}\) [$E_h$] $\downarrow$} & \multirow{2}{*}{$\epsilon$ [$E_h$] $\downarrow$} & \multirow{2}{*}{$\psi$ $\uparrow$} \\
      \cmidrule(lr){3-5}
      ~ & ~ & diagonal & non-diagonal & all & ~ & ~ \\
      \midrule
      \multirow{2}{*}{QH9-stable-id} & {\em MGAHam} & $(5.15 \pm 1.19) \times 10^{-2}$ & $(3.76 \pm 0.01) \times 10^{-2}$ & $(3.84 \pm 0.05) \times 10^{-2}$ & $(6.74 \pm 0.12) \times 10^{1}$ & $(4.87 \pm 1.75) \times 10^{-2}$ \\
      ~ & {\em MGAHam} (1D+3D) & \cellcolor{MyGreen4}$(3.23 \pm 0.33) \times 10^{-2}$ & \cellcolor{MyGreen4}$(2.30 \pm 0.36) \times 10^{-2}$ & \cellcolor{MyGreen4}$(2.54 \pm 0.56) \times 10^{-2}$ & \cellcolor{MyGreen4}$(2.11 \pm 2.23) \times 10^{1}$ & \cellcolor{MyGreen4}$(6.21 \pm 0.36) \times 10^{-2}$ \\
      \midrule
      \multirow{2}{*}{QH9-stable-ood} & {\em MGAHam} & $(4.67 \pm 0.33) \times 10^{-2}$ & $(3.15 \pm 0.51) \times 10^{-2}$ & $(3.58 \pm 0.49) \times 10^{-2}$ & $(6.41 \pm 0.38) \times 10^{1}$ & $(4.95 \pm 1.77) \times 10^{-2}$ \\
      ~ & {\em MGAHam} (1D+3D) & \cellcolor{MyGreen4}$(3.02 \pm 0.66) \times 10^{-2}$ & \cellcolor{MyGreen4}$(1.67 \pm 0.35) \times 10^{-2}$ & \cellcolor{MyGreen4}$(2.06 \pm 0.03) \times 10^{-2}$ & \cellcolor{MyGreen4}$(1.28 \pm 0.99) \times 10^{1}$ & \cellcolor{MyGreen4}$(5.76 \pm 0.84) \times 10^{-2}$ \\
      \midrule
      \multirow{2}{*}{QH9-dynamic-geo} & {\em MGAHam} & $(5.25 \pm 1.03) \times 10^{-2}$ & $(3.81 \pm 0.08) \times 10^{-2}$ & $(3.96 \pm 0.07) \times 10^{-2}$ & $(6.70 \pm 0.09) \times 10^{1}$ & $(4.76 \pm 1.56) \times 10^{-2}$ \\
      ~ & {\em MGAHam} (1D+3D) & \cellcolor{MyGreen4}$(3.46 \pm 0.92) \times 10^{-2}$ & \cellcolor{MyGreen4}$(2.18 \pm 1.10) \times 10^{-2}$ & \cellcolor{MyGreen4}$(2.50 \pm 0.77) \times 10^{-2}$ & \cellcolor{MyGreen4}$(1.96 \pm 1.69) \times 10^{1}$ & \cellcolor{MyGreen4}$(5.82 \pm 2.23) \times 10^{-2}$ \\
      \midrule
      \multirow{2}{*}{QH9-dynamic-mol} & {\em MGAHam} & $(5.06 \pm 0.88) \times 10^{-2}$ & $(3.90 \pm 0.04) \times 10^{-2}$ & $(3.97 \pm 0.46) \times 10^{-2}$ & $(6.98 \pm 0.68) \times 10^{1}$ & $(4.95 \pm 1.57) \times 10^{-2}$ \\
      ~ & {\em MGAHam} (1D+3D) & \cellcolor{MyGreen4}$(5.01 \pm 0.76) \times 10^{-2}$ & \cellcolor{MyGreen4}$(3.60 \pm 0.15) \times 10^{-2}$ & \cellcolor{MyGreen4}$(3.79 \pm 0.33) \times 10^{-1}$ & \cellcolor{MyGreen4}$(4.44 \pm 0.36) \times 10^{1}$ & \cellcolor{MyGreen4}$(5.07 \pm 1.01) \times 10^{-2}$ \\
      \midrule
      \multirow{2}{*}{QH-BM-id} & {\em MGAHam} & $(2.15 \pm 0.40) \times 10^{-1}$ & $(2.10 \pm 0.07) \times 10^{-1}$ & $(2.03 \pm 0.16) \times 10^{-1}$ & $(2.05 \pm 0.88) \times 10^{1}$ & \cellcolor{MyGreen4}$(5.22 \pm 0.57) \times 10^{-1}$ \\
      ~ & {\em MGAHam} (1D+3D) & \cellcolor{MyGreen4}$(2.13 \pm 0.81) \times 10^{-1}$ & \cellcolor{MyGreen4}$(1.98 \pm 0.75) \times 10^{-1}$ & \cellcolor{MyGreen4}$(1.86 \pm 0.70) \times 10^{-1}$ & \cellcolor{MyGreen4}$(1.99 \pm 0.70) \times 10^{1}$ & $(1.46 \pm 0.55) \times 10^{-1}$ \\
      \midrule
      \multirow{2}{*}{QH-BM-ood} & {\em MGAHam} & $(2.48 \pm 0.79) \times 10^{-1}$ & $(2.59 \pm 0.76) \times 10^{-1}$ & $(2.47 \pm 0.72) \times 10^{-1}$ & \cellcolor{MyGreen4}$(2.63 \pm 1.65) \times 10^{1}$ & \cellcolor{MyGreen4}$(1.33 \pm 0.39) \times 10^{-1}$ \\
      ~ & {\em MGAHam} (1D+3D) & \cellcolor{MyGreen4}$(2.63 \pm 0.99) \times 10^{-1}$ & \cellcolor{MyGreen4}$(2.52 \pm 0.95) \times 10^{-1}$ & \cellcolor{MyGreen4}$(2.42 \pm 0.91) \times 10^{-1}$ & $(5.76 \pm 2.18) \times 10^{1}$ & $(1.28 \pm 0.48) \times 10^{-1}$ \\
      \bottomrule
    \end{tabular}
  }
    \vspace{-0.2cm}
\end{table*}

A clear comparison of both prediction performance and computational efficiency is presented in Fig. \ref{fig:radar} (a), comparing \textit{MGAHam} against other algorithms. As illustrated, \textit{MGAHam} clearly surpasses QHNet and DEQHNet across all evaluated metrics, including Hamiltonian accuracy, orbital energies, occupied orbital coefficients, and computational efficiency. The exceptional computational efficiency, combined with the high prediction accuracy, makes it a powerful tool for AI-based high-throughput quantum chemistry.

In addition, as depicted in Fig. \ref{fig:mgaPara}, the parameter distribution within the framework demonstrates that a small geometric model ---  GNN (20.4 MB) guides a much larger language model --- SMILES-BERT (109.1 MB). The fact that the alignment layer requires very few parameters indicates that the features extracted by the GNN are highly representative. This GNN defines a compact geometric space that provides a clear training target for the BERT model. Consequently, the framework can predict the properties by only using SMILES strings, skipping the time-consuming 3D geometry acquisition step. These results confirm that a lightweight alignment approach is sufficient to link 1D sequences with 3D geometric information.

 \subsection{Discussion on the Scalability}

While our proposed \textit{MGAHam} framework has demonstrated promising accuracy and efficiency in predicting Hamiltonians from SMILES strings, it's crucial to acknowledge an inherent issue: a single molecule often has multiple conformations. These conformations could lead to different potential energies, diverse behaviors, and different Hamiltonians. Therefore, while SMILES-only prediction excels in high-throughput screening, molecular geometries are still required for certain scenarios, e.g., accurately predicting the Hamiltonian of different conformations. In this section, we showcase \textit{MGAHam}'s capability in supporting both SMILES strings and molecular geometries for more accurate Hamiltonian predictions.

To achieve this capability, \textit{MGAHam} (1D+3D) employs a feature fusion strategy. During the inference stage, geometric embeddings are extracted using QHNet. These geometrically derived features are then summed with the token embeddings obtained from a frozen SMILES-BERT. The resulting combined embeddings are subsequently fed into the prediction head, allowing the model to leverage the complementary information from both modalities.

We first conducted experiments on the MD17 dataset, which consists of 4 molecules with thousands of conformations per molecule. Table \ref{tab:md17} presents the MAEs for Hamiltonian $\mathbf{H}$, total energy ($\epsilon$), and the cosine similarity ($\psi$) on the MD17 dataset. From the results, \textit{MGAHam} (1D+3D) consistently achieves the best performance in most cases. Specifically, on the Water molecule, \textit{MGAHam} (1D+3D) achieves an $\mathbf{H}$ error of $1.73 \times 10^{-2}E_h$ and an $\epsilon$ error of $0.142 E_h$, significantly outperforming all other baselines, including SchNOrb, PhiSNet, QHNet, and DEQHNet. For Ethanol, Malonaldehyde, and Uracil, where \textit{MGAHam} (1D+3D) frequently secures the good performance on the Hamiltonian accuracy, underscoring its superior accuracy in capturing the subtle nuances of Hamiltonian changes due to molecular conformations. 
While strong baselines like SE(3)-Transformer and GemNet show competitive results, they generally fall short of the overall predictive accuracy achieved by \textit{MGAHam}.

% . For instance, on the Malonaldehyde dataset, GemNet achieves a competitive $\mathbf{H}$ MAE of $1.60  \times 10^{-2} E_h$, which is slightly higher than the value of \textit{MGAHam} (${1.36  \times 10^{-2} E_h}$). Furthermore, although SE(3)-Transformer achieves the best $\epsilon$ and $\psi$ scores in certain specific cases (Ethanol $\epsilon$ and Uracil $\epsilon$ and $\psi$), it fails to match the performance of \textit{MGAHam} on the critical $\mathbf{H}$ metric for these same molecules.

To verify the benefits of combining both molecular geometries and SMILES strings, we conducted further experiments on the QH9 dataset. As Table \ref{tab:1dvs3d} shows, compared to \textit{MGAHam}, \textit{MGAHam} (1D+3D) leads to substantial improvements. Specifically, a 38.3\% improvement for the Hamiltonian matrix ($\mathbf{H}$), 286.5\% for energy ($\epsilon$), and 150.9\% for cosine similarity ($\psi$).
When compared to the SMILES-BERT model and QHNet (as detailed in Table \ref{tab:qh9_without_deqhnet}), \textit{MGAHam} (1D+3D) also shows significant gains in prediction accuracy, 35.5\% for $\mathbf{H}$, 85.6\% for $\epsilon$, and 140.2\% for $\psi$. 
Overall, as illustrated in Fig. \ref{fig:radar} (b), compared with MGAHam, MGAHam (1D+3D) sacrifices computational efficiency, but it achieves higher accuracy in quantum property prediction.
 
% Beyond prediction accuracy, we also investigate the computational efficiency when using both molecular geometries and SMILES strings. As illustrated in Table \ref{tab:accel}, the acceleration rates relative to DFT calculations are still stark. Specifically, the computation cost of \textit{MGAHam} for 1,000 molecules on QH9 is about 6 minutes and for QH-BM, about 48 minutes.
% \textit{MGAHam} is approximately 99.42$\times$ faster than DFT, while on the QH-BM dataset, it reaches 100.05$\times$ acceleration. Compared to other GNN baselines, such as QHNet and DEQHNet, \textit{MGAHam} operates 3-5 times faster. 

\subsection{Ablation Study and Parameter Analysis}

We conducted a series of ablation studies to systematically evaluate the contribution of each key component of our model. Specifically, we test the resulting performance changes by individually removing the multimodal alignment module and the weakly supervised fine-tuning module. Our results on the QH-BM dataset, using SMILES-BERT as the backbone (detailed in Supplemental Material, confirmed that both components are vital. The full \textit{MGAHam} model achieved the lowest overall $\mathbf{H}$ MAE. Using only pre-training or only fine-tuning on this backbone provided partial improvements, yielding $\mathbf{H}$ MAEs of $0.244 E_h$ and $0.240 E_h$, respectively. The improved performance is attributed to the multimodal alignment, allowing the model to capture critical spatial information from SMILES, and the weak supervision addressing the low-data issue to enhance robustness.

To further verify the scalability of our \textit{MGAHam}, we assess the performance when using another molecular language BERT --- ChemBERTa \cite{chithrananda2020chemberta} and GNN backbone --- DEQHNet. Furthermore, we investigated the impact of our loss function's design on the final results. We performed a detailed parameter analysis to examine how varying the ratio of the diagonal and off-diagonal components of the loss function affects model performance.
The detailed results for both the ablation studies and parameter analysis are presented in the Supplemental Material.

\subsection{A Case Study: Screening of Electrolyte Formulation}

\begin{figure*}[!ht]
	\centering
	\includegraphics[width=0.99\linewidth]{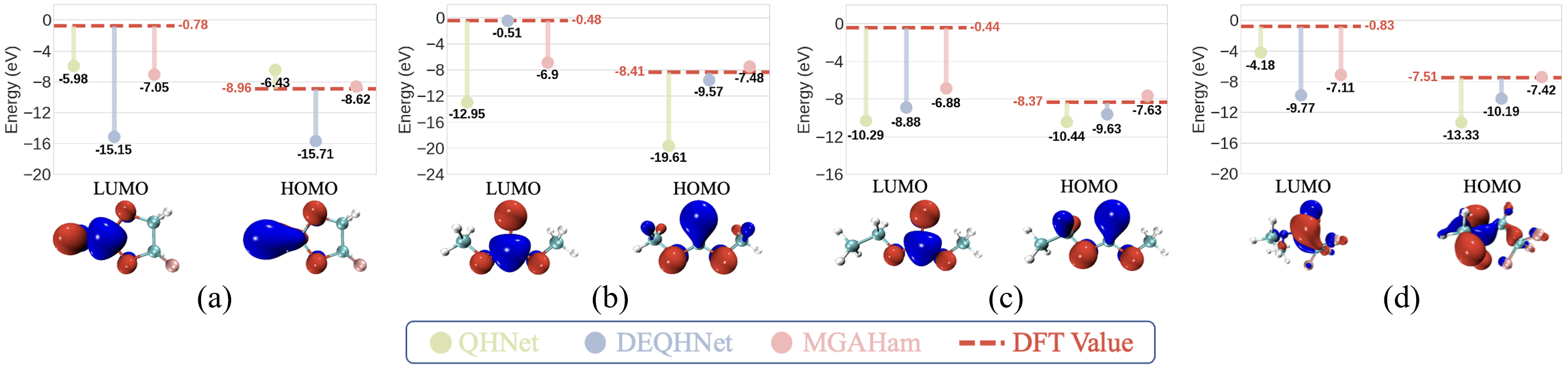}
	\caption{The derived and DFT-calculated HOMO and LUMO values of FEC (Fluoroethylene Carbonate), DMC (Dimethyl Carbonate), EMC (Ethyl Methyl Carbonate), and FDMA (Fluorodimethoxyethane).}
	\label{fig:HOMO&LUMO_5moles}
    \vspace{-0.4cm}
    
\end{figure*}

Lithium-ion batteries are the backbone of modern portable electronics and electric vehicles. Continuous advancements in their performance are fundamentally tied to the meticulous design and engineering of electrolytes. Traditionally, the process of screening and optimizing electrolyte formulations---identifying the most suitable solvents and salts---has been computationally intensive and resource-demanding. Here, we leveraged our proposed {\em MGAHam} to screen potential electrolyte formulations.

The solid electrolyte interphase (SEI) plays a critical role in lithium-ion battery performance, particularly in maintaining battery stability. SEI formation initiates when the lithium-ion-rich solvent layer near the lithium metal surface undergoes reduction by accumulating electrons. To control the SEI's composition, it is crucial that the preferred interfacial species possess a higher reduction potential compared to the main solvents. From a molecular orbital perspective, this directly translates to a higher electron affinity or, equivalently, a lower LUMO energy. Molecules with lower LUMO energies are more prone to accepting electrons and thus decompose preferentially to form a stable SEI.

Our screening process, as illustrated in Fig. \ref{fig:HOMO&LUMO_5moles}, involved investigating four different solvents. Crucially, {\em MGAHam}'s predictions for LUMO energies highlighted fluorinated dimethylacetamide (FDMA) as the solvent with the lowest LUMO energy among all candidates. This strong indication of high electron affinity suggests that FDMA is thermodynamically favored to decompose first, thereby playing a dominant role in directing the SEI formation.

However, FDMA itself presents a relatively high predicted HOMO level, which could potentially compromise the electrolyte's high-voltage stability when paired with the cathode. To counteract this, incorporating fluoroethylene carbonate (FEC) as a co-solvent becomes a strategic necessity. FEC is predicted to have the lowest HOMO level among widely used solvents. Its inclusion is critical because it facilitates the formation of lithium fluoride within the SEI, a known mechanism that significantly enhances the electrolyte's high-voltage stability towards the cathode, as supported by previous research \cite{fan2018non}. The synergistic combination of FDMA for controlled SEI initiation and FEC for high-voltage stability forms a robust solvent system.

Regarding the choice of lithium salts, while LiNO$_3$ is a widely recognized N-donating additive known for its benefits in SEI formation \cite{jozwiuk2016critical}, its practical insolubility in carbonate solvents renders it unfeasible for many common electrolyte formulations.
When considering LiPF$_6$, a prevalent lithium salt, a significant challenge arises: FEC, despite its benefits as a co-solvent, exhibits thermal instability in LiPF$_6$-based electrolytes. This instability triggers a detrimental cascade, leading to the generation of unwanted hydrofluoric acid and various other acids. These acidic byproducts are highly corrosive, causing substantial dissolution of transition metal ions from the cathode into the electrolyte. Such dissolution directly contributes to the severe degradation of cathode materials \cite{han2019scavenging}, drastically shortening battery lifespan.

In contrast, other frequently used salts like LiTFSI and LiFSI share a similar electrochemical window, offering comparable performance profiles. However, LiTFSI stands out for its superior stability towards lithium metal, a critical advantage attributed to its robust $-$CF$_3$ group \cite{han2019scavenging}. This enhanced stability makes LiTFSI a more reliable choice for long-term cycling performance in lithium-metal batteries.
Based on the predictions from our {\em MGAHam} and the analysis of solvent and salt properties, we deduced that an electrolyte comprising LiTFSI in an FDMA/FEC mixed solvent represents an advisable formulation. 
This formulation has been proven to exhibit favorable electrochemical stability and high reactivity~\cite{wang2020interface}.

\section{Conclusions}

In this work, we introduce \textit{MGAHam}, a novel and effective deep learning framework for predicting molecular Hamiltonians. Our method addresses a critical bottleneck often encountered by deep learning surrogates: the substantial requirement for expensive molecular geometries and large-scale Hamiltonian matrices as training data. \textit{MGAHam} circumvents these limitations by leveraging a multimodal alignment strategy coupled with modality compensation. This approach establishes a robust intrinsic relationship between SMILES strings and 3D molecular structures, effectively eliminating the need for precise geometries during inference. Furthermore, to enhance generalizability in data-constrained scenarios, \textit{MGAHam} incorporates a mask-based weakly supervised fine-tuning stage, enabling high-performance adaptation even with highly limited Hamiltonian data. Our theoretical analysis demonstrates that the generalization error arising from the absence of explicit 3D geometry remains bounded within the proposed framework. Extensive experimental validations---spanning four benchmark datasets under both in-distribution and out-of-distribution settings, as well as an electrolyte formulation screening case study---demonstrate \textit{MGAHam}'s superior performance in terms of accuracy and computational efficiency. This innovative framework establishes a promising foundation for AI-driven high-throughput quantum chemistry.

\bibliography{example_paper}

\bibliographystyle{IEEEtran}

\end{document}